%% file: main.tex
\documentclass[journal]{vgtc}                     

\usepackage{times}                     

\usepackage{graphicx}
\usepackage{array}
\usepackage{makecell}
\usepackage{amsmath}
\usepackage{amsfonts}
\usepackage{xcolor}
\usepackage{wrapfig}
\onlineid{0}

\vgtccategory{Research}

\vgtcpapertype{algorithm / technique}

\title{AnyTalk: Speech Animation for Arbitrary Characters \\ Leveraging a Video Generation Model}

\author{%
 \authororcid{Kwan Yun*}{0000-0002-0205-5203},
 \authororcid{Serin Yoon*}{0009-0008-7464-9968},
 \authororcid{Sunjin Jung}{0000-0001-6427-6258},
 \authororcid{Jung Eun Yoo}{0000-0002-6276-452X},
 \authororcid{Inyup Lee}{0009-0003-6014-6132}, and 
 \authororcid{Junyong Noh}{0000-0003-1925-3326}
}

\authorfooter{
    Kwan Yun (yunandy@kaist.ac.kr), 
    Serin Yoon (serinyoon@kaist.ac.kr), 
    Sunjin Jung (sunjin@sungshin.ac.kr), 
    Jung Eun Yoo (jey.kaist@gmail.com), 
    Inyup Lee (leeinyup123@kaist.ac.kr), 
    Junyong Noh (junyongnoh@kaist.ac.kr).  
    \\ * Contributed equally to this work
}
\newcommand{\fixq}[1]{\textcolor{black}{#1}}

\teaser{
  \includegraphics[width=\textwidth]{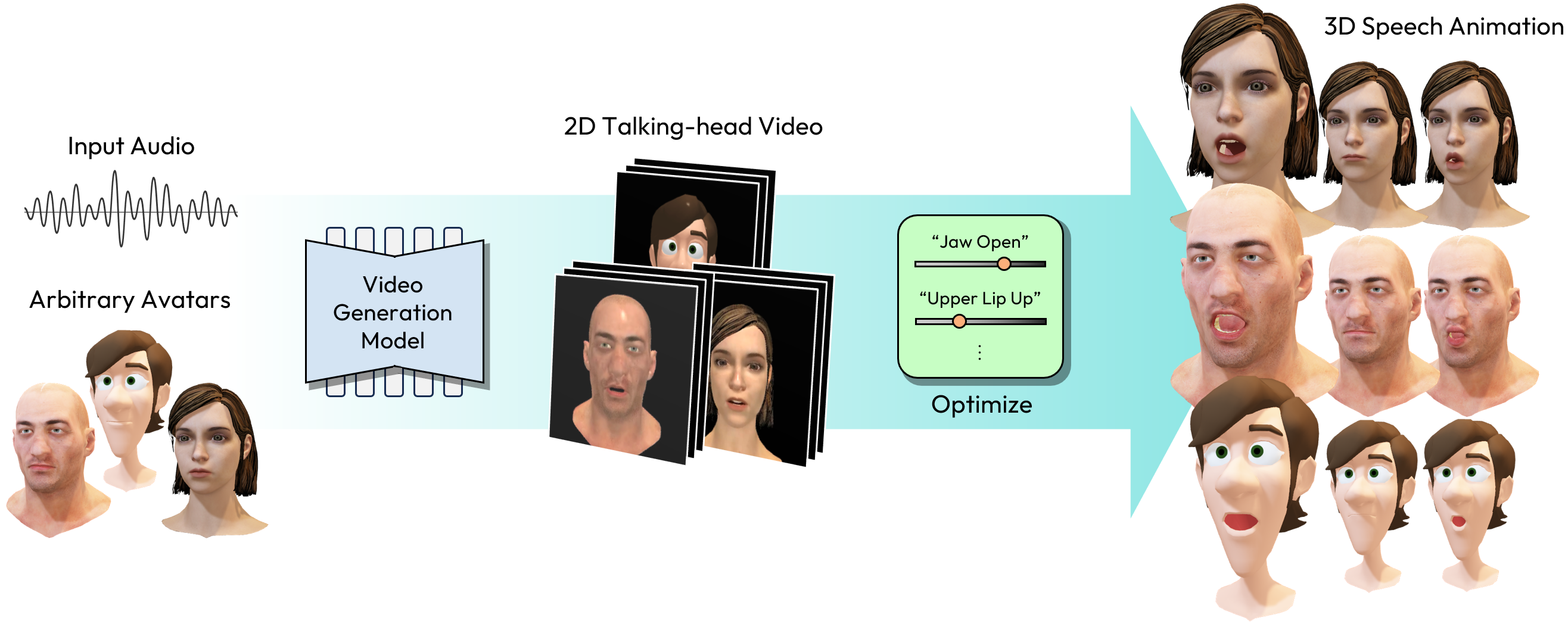}

  \refstepcounter{figure}
  \label{fig:teaser}

  \vspace{\abovecaptionskip}
  {\scriptsize\sffamily
  \textbf{Fig. \thefigure.}
  AnyTalk can perform audio-driven 3D facial animation using an arbitrary character with blendshapes,
  without requiring any animation data to train.
  \par}
  \vspace{\belowcaptionskip}
}

\input{sec/0_abstract}
\keywords{Audio-driven animation, facial animation, diffusion model.}

\begin{document}


\firstsection{Introduction}

\maketitle
  
\input{sec/1_intro}

\input{sec/2_related_work}

\input{sec/3_method}
\input{sec/4_experiments}

\input{sec/5_application}

\input{sec/6_conclusion}


\bibliographystyle{abbrv-doi}

\bibliography{main}

\begin{minipage}[h]{\columnwidth}
\begin{wrapfigure}{l}{25mm}\vspace{-5mm} 
    \includegraphics[width=25mm,keepaspectratio]{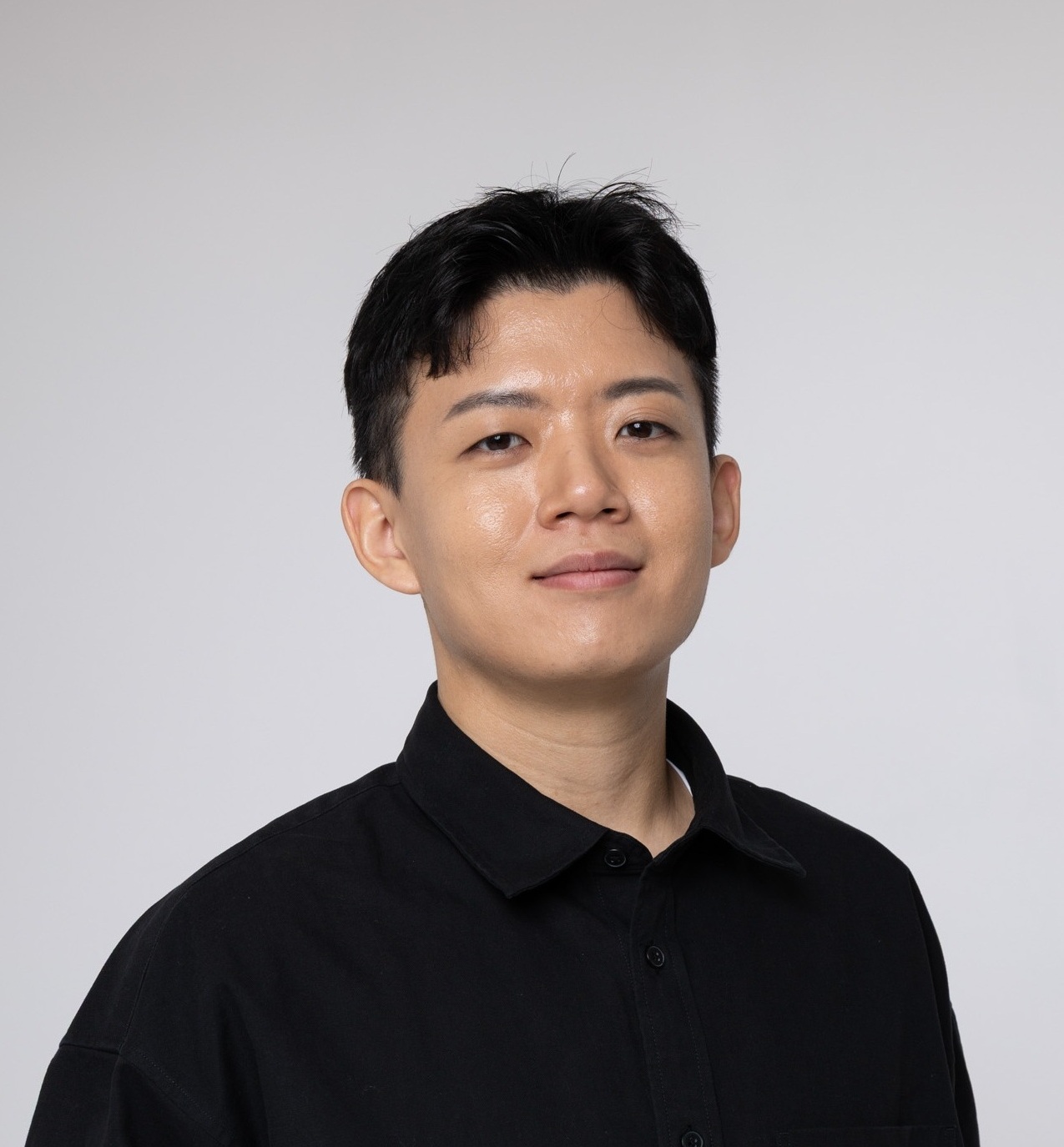}
\end{wrapfigure}
\textbf{Kwan Yun} is a Ph.D student in the Graduate School of Culture Technology at Korea Advanced Institute of Science and Technology (KAIST). He earned his M.S. from KAIST in 2024. His research focuses on leveraging generative models for face and character stylization, animation, and editing.
\end{minipage}
\vspace{5mm}

\begin{minipage}[h]{\columnwidth}
\begin{wrapfigure}{l}{25mm}\vspace{-5mm} 
    \includegraphics[width=25mm,keepaspectratio]{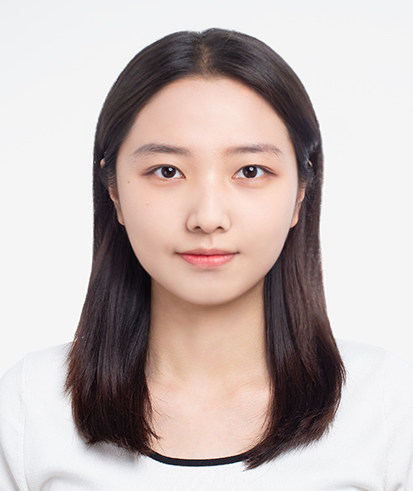}
\end{wrapfigure}
\textbf{Serin Yoon} is currently a visiting researcher at the DGP Lab, University of Toronto. She received her master's degree from the Graduate School of Culture Technology at KAIST in 2026, and her bachelor's degree in Computer Science Education from SKKU in 2024. Her research interests include computer graphics, computer vision, and 3D animation.
\end{minipage}
\vspace{5mm}

\begin{minipage}[h]{\columnwidth}
\begin{wrapfigure}{l}{25mm}\vspace{-5mm} 
    \includegraphics[width=25mm,keepaspectratio]{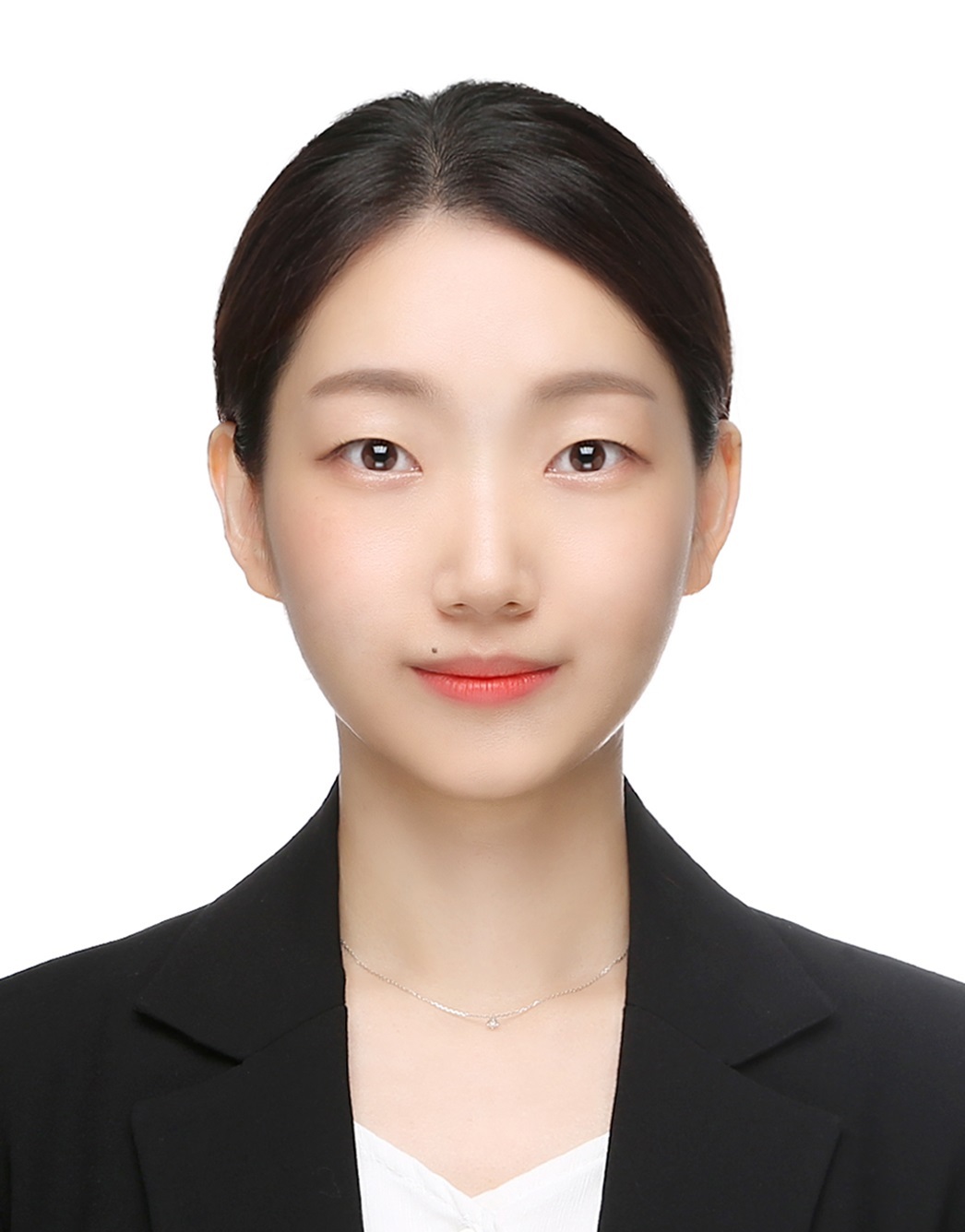}
\end{wrapfigure}
\textbf{Sunjin Jung} is an Assistant Professor in the Department of Computer Engineering at Sungshin Women’s University. She received her M.S. and Ph.D. degrees from the Korea Advanced Institute of Science and Technology (KAIST). Her research interests include computer graphics and character animation.
\end{minipage}
\vspace{5mm}

\begin{minipage}[h]{\columnwidth}
\begin{wrapfigure}{l}{25mm}\vspace{-5mm} 
    \includegraphics[width=25mm,keepaspectratio]{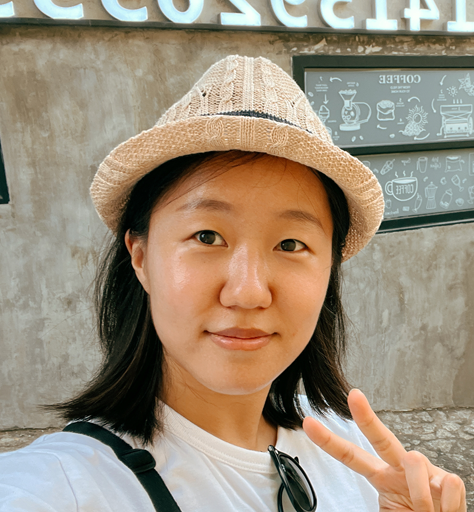}
\end{wrapfigure}
\textbf{Jung Eun Yoo} is an R\&D engineer who received a Ph.D. in 2025 from the Graduate School of Culture Technology at the Korea Advanced Institute of Science and Technology (KAIST). Her research focuses on applying AI within creative workflows to streamline and simplify the content creation process..
\end{minipage}
\vspace{5mm}

\begin{minipage}[h]{\columnwidth}
\begin{wrapfigure}{l}{25mm}\vspace{-5mm} 
    \includegraphics[width=25mm,keepaspectratio]{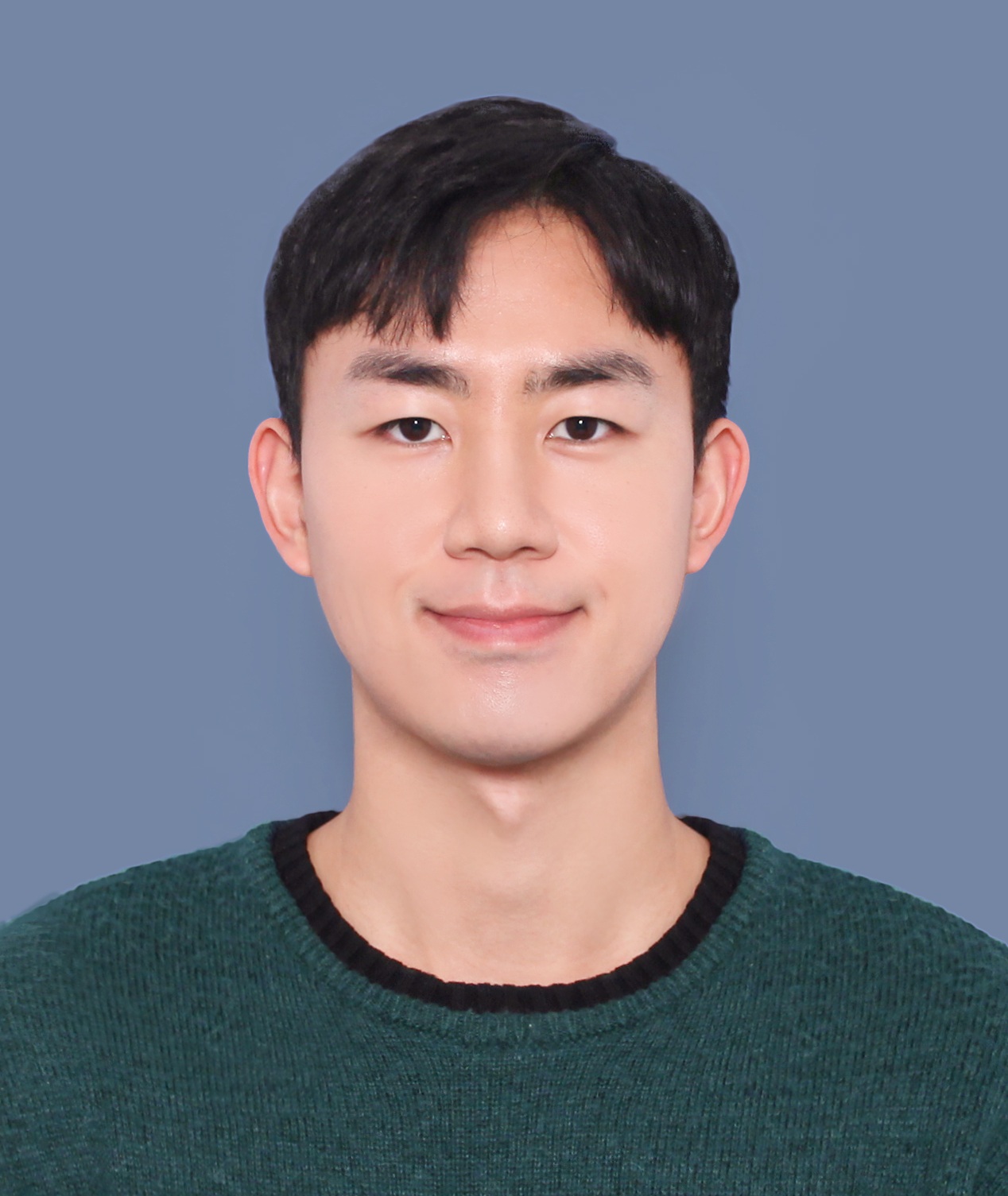}
\end{wrapfigure}
\textbf{Inyup Lee} is a Ph.D student in the Graduate School of Culture Technology at Korea Advanced Institute of Science and Technology (KAIST). He earned his M.S. from KAIST in 2025. His research focuses on facial animation, facial animation editing and facial reconstruction.
\end{minipage}
\vspace{5mm}

\begin{minipage}[h]{\columnwidth}
\begin{wrapfigure}{l}{25mm} 
    \includegraphics[width=25mm,keepaspectratio]{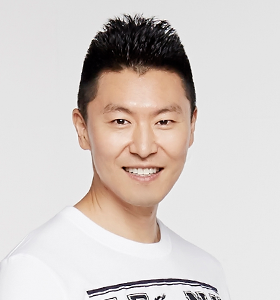}
\end{wrapfigure}
\textbf{Junyong Noh} is a Professor in the Graduate
School of Culture Technology (GSCT) at KAIST. He earned the Ph.D. degree in computer science (2002), the master’s degree in computer engineering (1996), and the bachelor’s degree in electrical engineering (1994) all from the University of Southern California (USC). His current research focus includes facial/character animation, virtual/augmented reality, image/video manipulation for immersive and interactive experiences. Prior to his academic career, he was a graphics scientist at a Hollywood visual effects company, Rhythm and Hues Studios. He held the title of KAIST chair professorship (2011) and received a technical innovation award from KAIST (2011). A research result, ScreenX, was selected as one of ten most representative research outcomes from KAIST (2013) and later successfully commercialized by the leading movie theater chain in Korea, CGV. Recently, he received the research innovation award at the 50th anniversary of KAIST.
\end{minipage}
\vspace{5mm}

\end{document}


\maketitle

\input{sec/99_supplementary}

\bibliographystyle{abbrv-doi}
\bibliography{main}

%% file: sec/0_abstract.tex
\abstract{
We present AnyTalk, a novel method for generating 3D speech animations for arbitrary characters without requiring any animation data. While existing audio-driven 3D speech animation methods rely on character-specific training data or laborious rigging/re-meshing, AnyTalk circumvents these limitations by leveraging recent video diffusion models trained on extensive video datasets. We first adapt a pre-trained video diffusion model to a target character through our Character-specific Fine-tuning (\textit{CsF}) technique. By fine-tuning on rendered images of the 3D character paired with zeroed-out audio embeddings (representing “no motion”), we eliminate the need for animation data while preserving the motion prior of large-scale video diffusion model. We then uplift the resulting talking-head video into a 3D speech animation by estimating blendshape parameters through a proposed optimization process. AnyTalk enables lip-synced animations across diverse face meshes and blendshape configurations, significantly reducing manual effort and data requirements. We further enhance usability by distilling AnyTalk into a streamlined network, $\text{AnyTalk}_{RT}$, thereby enabling real-time performance. By leveraging talking-head video generation, our method broadens access to audio-driven speech animation technology for arbitrary characters. The code is publicly available at \href{https://serin-yoon.github.io/projects/anytalk/}{AnyTalk}.
}

%% file: sec/1_intro.tex
\label{sec:intro}
The increasing demand for realistic 3D speech animation across various media platforms such as games and VR/AR highlights the necessity of advanced techniques in the animation industry. 
Manually keyframing a character's lip movements to synchronize with audio has been a popular choice. However, it is both time-consuming and requires significant expertise, making it impractical for non-professionals and cumbersome even for experienced artists.

To address this challenge, deep-learning-based audio-driven speech animation methods~\cite{fan2022faceformer,xing2023codetalker,cudeiro2019capture,jung2024speed,richard2021meshtalk} have been proposed, substantially reducing the burden of manual keyframing. Nevertheless, applying these advanced techniques to real-world characters remains challenging. This is largely because most existing audio-driven 3D facial animation methods must be trained with paired audio–3D data for each unique mesh structure or blendshape configuration. Consequently, rigging artists must either re-rig their characters to match existing blendshapes or gather training data for every mesh structure. This requirement can pose a significant barrier to the widespread adoption of audio-driven speech animation technique, because re-rigging characters or acquiring matched training data for each unique mesh structure is difficult for independent developers or small studios with limited resources.


Meanwhile, recent advancements in diffusion-based talking-head video generation models~\cite{xu2024hallo,cui2025hallo,xu2024vasa,wang2024v,wei2024aniportrait} have successfully produced realistic talking-head videos. This achievement is largely attributed to training large diffusion models~\cite{ho2020denoising} on extensive and diverse video datasets spanning hundreds to thousands of hours. As a result, these models can generate lip-synced videos from unseen characters, which has been difficult for most 3D speech animation methods. \fixq{By utilizing these high-quality 2D outputs as an intermediate representation, our method effectively bridges the gap between the rich motion priors of 2D generative models and the practical requirements of 3D facial animation.}

Building on this success, we propose AnyTalk, a novel method that leverages a video generation model to produce 3D speech animation for arbitrary avatars as shown in Figure~\ref{fig:teaser}. Our method follows a two-step process: (1) \textbf{Talking-head Video Generation:} Given a 3D character, we render an image of the character and generate a speech-synchronized talking-head video using a video generation model. (2) \textbf{Blendshape Optimization:} We estimate blendshape parameters from the generated video using an optimization process to create accurate 3D speech animation.
This sequential approach utilizes the rich motion prior and generalizability of diffusion-based talking-head video generation models, enabling speech animation for arbitrary 3D characters.

However, directly applying pre-trained video generation models produces significant visual artifacts and poor character fidelity. Although the output videos synchronize well with the audio, they may not conform to the 3D character's range of motion or correctly depict previously unseen facial attributes. This discrepancy between the generated video and the character’s inherent motion capabilities not only compromises the visual fidelity but also complicates the downstream blendshape optimization, making it more difficult to produce precise 3D speech animation.


To overcome this visual mismatch, we propose Character-specific Fine-tuning ($CsF$) without requiring video data for audio-driven video generation models. Instead of relying on ground-truth audio-visual pairs, we use the rendered images of the character paired with zeroed-out audio embeddings as input. Because the audio input is the driving source of motion, zeroing it effectively pairs each rendered still image with “no motion signal” during training while allowing a non-zero speech signal at inference. To further preserve motion priors, we freeze the temporal module while updating the spatial module of the diffusion model, ensuring that the visual information of the generated video aligns well with the rendered images of the character while the trained motion prior is not modified.

To obtain blendshape parameters from the generated video, we propose a landmark matching optimization technique. Specifically, we first estimate landmarks from the rendered image, cast rays to identify the corresponding vertices, and then optimize the blendshape parameters so that these vertices align with the landmarks of the target video when rendered. In addition, during landmark-based optimization, we select talk invariant landmarks, estimate homography, and warp talk-related landmarks accordingly to further enhance landmark alignment.

By integrating the proposed techniques, AnyTalk successfully produces 3D speech animation for arbitrary characters without requiring any animation data for training. In summary, our contributions are as follows:
\begin{itemize}
    \item AnyTalk is the first 3D speech animation method for arbitrary characters that does not rely on \fixq{3D animation data} for training. This is achieved by utilizing a 2D talking-head generation model and uplifting the generated video into 3D animation.
    \item $CsF$ is a novel fine-tuning method for talking-head generation models that enables the generated video to match a specific character without requiring any video data.
    \item Our homography-based warping, driven by talk-invariant landmark selection, ensures precise alignment between the rendered images of arbitrary characters and the generated video. This alignment enables accurate estimation of blendshape parameters for the characters from generated video.

    \item We develop a real-time variant, $\text{AnyTalk}_{RT}$, using model distillation with feature matching and reconstruction losses, achieving an inference speed of 110 FPS for real-time applications.
\end{itemize}

%% file: sec/2_related_work.tex
\section{Related Work}
\subsection{Audio-driven Talking-head Video Generation}
Generating 2D talking-head videos is a long-standing and rapidly advancing field with wide usability~\cite{suwajanakorn2017obama,lu2021live,vougioukas2020realistic,prajwal2020wav2lip,zhou2021pcavs, wang2021audio2head, seo2022spv,wang2022one,yang2025infinitetalk,kong2025let}. Synthesizing 2D facial animation from audio typically utilizes various image representations, such as facial landmarks~\cite{zhou2020makeittalk,ye2023geneface,ji2021evp}, 3D Morphable Model parameters~\cite{zhang2023sadtalker}, depth maps~\cite{hong2022depth}, and semantic maps~\cite{liu2022semantic} to facilitate the transformation of speech into 2D facial animations. More recently, diffusion and flow based methods~\cite{shen2023difftalk,xu2024vasa,tian2024emo,xu2024hallo,wang2024v,cui2025hallo,lin2025omnihuman} have been proposed. These methods advanced the realism of audio-driven talking-head video generation by utilizing larger networks and more extensive data. Therefore, we fine-tune these models to make them applicable to a target 3D character for speech animation.


\subsection{Audio-driven Speech Animation}
\fixq{Recently, significant efforts have been made in audio-driven 3D facial animation~\cite{aneja2023facetalk,thambiraja20233diface,richard2021meshtalk,karras2017audio,thambiraja2023imitator,danvevcek2023emotional,zhou2018visemenet,pan2022vocal,pan2025vasa,chai2022personalized,sun2024diffposetalk,fan2024unitalker,li2025wav2sem,gu2025diverse,liu2021geometry}.} VOCA~\cite{cudeiro2019capture} approaches the transformation from audio to animation as a regression problem, using paired audio and 3D face animation data to train the regressor. FaceFormer~\cite{fan2022faceformer} utilizes transformers to handle the long-term dependencies while CodeTalker~\cite{xing2023codetalker} utilizes quantized latent space to create natural motion. More recently, diffusion-based methods~\cite{shen2023difftalk,ma2024diffspeaker,stan2023facediffuser, song2024expressive} have been proposed to add realism and diversity to the generated video. \fixq{Another line of research extends from facial animation to generating both facial expressions and gestures guided by audio~\cite{zhang2025echomask,zhang2025semtalk}}. All the aforementioned methods were trained in a supervised manner, which requires paired audio-3D data.

\subsection{Audio-driven Speech Animation for Diverse Characters}

To address the large dataset requirement for each character, several methods have been proposed that focus on single-identity-based approaches. \textit{taylor et al.}~\cite{taylor2017deep} proposed an audio-driven speech animation method with retargeting, assisted by transcription. Similarly, \textit{yang et al.}~\cite{yang2023semi} map single-identity audio and a single character into the same embedding space to generate a character motion in a semi-supervised manner. While both approaches require only audio-visual data for a single identity, they have limited applicability to different audio sources. For example, \textit{taylor et al.}~\cite{taylor2017deep} requires manual specification of the correspondence for AAM shape components and transcriptions, making the process labor-intensive. On the other hand, \textit{yang et al.}~\cite{yang2023semi} is constrained to a single source identity and cannot generalize to other characters or audio sources.

Most recently, ScanTalk~\cite{nocentini2025scantalk} was proposed, which utilizes large paired audio-scan data across various mesh structures. This system marks the first 3D talking-head generator capable of handling diverse mesh structures using the DiffusionNet~\cite{sharp2022diffusionnet} architecture. However, ScanTalk was trained on meshes that were manually aligned, resampled, and modified to match the properties of VOCASET~\cite{cudeiro2019capture}. As a result, the model's performance varies significantly depending on how well a given mesh aligns with the training dataset, making it challenging to generate accurate speech animations for in-the-wild characters. In contrast, our approach generates 2D motion and uplifts it to 3D, without being constrained to any mesh structure or requiring mesh modifications.

\subsection{Fine-tuning Video Diffusion Model}
Fine-tuning video diffusion models for a specific character has become an active research area. Most previous personalization methods have either replaced the original layer with a personalized one~\cite{liew2023magicedit,guo2023animatediff} or masked out the temporal module and applied existing image-personalization techniques to the spatial module~\cite{molad2023dreamix}. More recently, Still-Moving~\cite{chefer2024still} proposed a method to fine-tune a general text-to-video diffusion model using a binary motion signal. Most of these methods aim to personalize the video diffusion model for a specific character, but not for a specific scene. On the other hand, AnyMoLe~\cite{yun2025anymole} adapts the model to both the character and the rendering scene using only a few seconds of 3D character motion. This is achieved through rectification training of an image-to-video diffusion model for the motion in-betweening task, using a few seconds of ground-truth motion. In fact, no method has personalized an audio-driven video diffusion model without relying on paired data. Our method is the first to personalize an audio-driven talking-head animation model to a target character without using any motion or audio data, employing a zeroed-out audio embedding as a stop-moving signal to disentangle audio from motion.

%% file: sec/3_method.tex
\section{Method}

At its core, our method consists of two main stages. In the first stage, we generate videos using an audio-driven video diffusion model. In the second stage, we estimate blendshape parameters from the generated video frames through an optimization process. \fixq{Because our motion generation process relies on blendshape estimation, the input character must be rigged.} The first stage generates the target motion, while the second stage produces the 3D animation based on the target motion. Each stage will be described in detail in the following subsections.

\input{figure_tex/03_00_baseline}

\subsection{Talking-head Video Generation Model}
We adopt a pre-trained Hallo~\cite{xu2024hallo}, a diffusion-based audio-driven video generation model which was trained on a large-scale human speech video dataset, as our baseline $D_{src}$. $D_{src}$ accepts a source image and driving audio as inputs and generates a corresponding video that matches both the source appearance and the audio content. A key feature of $D_{src}$ is its hierarchical audio processing via three distinct modules: pose, expression, and lip residual attention. This design enables the use of control weights $w_{pose}$, $w_{exp}$, and $w_{lip}$, as shown in Figure~\ref{fig:base_hallo}, to modulate motion dynamics. This controllability allows head motions to remain stable while the lips move dynamically, thereby enabling a second stage of video-based blendshape optimization with improved stability.

\subsection{Character-specific Fine-tuning without Speech Animation Data} \label{sec:csf}

Although recent audio-driven talking-head video generation models, including $D_{src}$, can be applied to a wide range of identities and audio inputs, we observed a mismatch between the generated videos and the rendered images of the 3D character. This mismatch hinders a direct application of these models to our framework because blendshape parameter estimation for animation relies on comparing the rendered images of the character with the generated video. The mismatch typically manifests in two forms: (1) undesired head motion and (2) input inconsistency. These issues stem from biases in the training data of $D_{src}$, which consists of natural human videos—captured in real-world settings, unstylized, and exhibiting natural head motion, rather than rendered scenes. Figure~\ref{fig:domain_gap} shows examples of these limitations suffered by recent methods, including Hallo~\cite{xu2024hallo}, AniTalker~\cite{liu2024anitalker}, EchoMimic~\cite{chen2025echomimic}, and MEMO~\cite{zheng2024memo}. Even though each method can generate a face that aligns with the audio, all four models suffer from visual inconsistencies. Hallo generates unnatural clothing and additional chin that do not exist in the rendered character; AniTalker shifts the target character toward a generic human-like appearance; EchoMimic produces close-up faces or exhibits black spots; and MEMO generates characters in unnatural mouth and teeth.

\input{figure_tex/03_01_domain_gap}
\input{figure_tex/03_02_training}

To address this, we introduce Character-specific Fine-tuning ($CsF$). $CsF$ adapts a general talking-head video generation model $D_{src}$ into a personalized model $D_{CsF}$ using only a target character. To achieve this transformation, we activate each blendshape parameter individually and render a frontal image for each, then duplicate these images to create a still image video as shown in Figure~\ref{fig:training}. Here, $R$ denotes the rendering function. Naively training the model to output a duplicated version of the input image given audio would cause it to forget how to animate effectively. For instance, if random speech is paired with still images during training, the model will produce a static video whenever it encounters similar audio, thereby losing its ability to animate. To prevent this, we zero out the audio encoder's output (audio embedding) during training. This zeroed-out audio embedding simulates a “no-motion” input, disentangling the no-motion signal from the general speech signal (i.e., motion). As a result, the model reproduces the original image when no motion signal is present, while retaining its capability to generate speech-driven motion when a valid audio signal is provided.

Because the purpose of $CsF$ is to learn the spatial features of the target character while preserving the original audio understanding, motion prior, and target reference understanding, we freeze all attention layers, including audio attention, motion attention, and the layers of ReferenceNet~\cite{hu2024animate}. Consequently, only the spatial residual network of denoising UNet is trained to accurately reproduce the visual appearance of the target character by predicting noise $\epsilon$. Formally, the training process is expressed as follows:
\begin{equation} 
L = \mathbb{E}_{\mathcal{E}(I), C_{zero},\epsilon\sim\mathcal{N}(0,1),t}\left[\lVert \epsilon-D_{CsF}(z_t, t, C_{zero}) \rVert_{2}^{2} \right]. 
\end{equation} 
Here, $t$ is uniformly sampled from diffusion timesteps $\{1,...,T\}$, $C_{zero}$ is zeroed-out audio embedding, and $z_t$ is noisy latent variable. Note that the control weights $w_{pose}$, $w_{exp}$, and $w_{lip}$ are not applied during the training.

\subsection{Video Inference}\label{sec:inference}
With $D_{CsF}$, a video for the target character can be generated within the domain of the target character. While training was performed using a zero motion condition, we provide general (non-zero) speech audio for inference to generate dynamic motions. We re-weight these attention mechanisms with scaling factors $w_{pose}$ = 0, $w_{exp}$ = 1, and $w_{lip}$ = 2 during inference as shown in Figure~\ref{fig:inference}. This ensures that $D_{CsF}$ produces speech with dynamic lip motion while maintaining a static head pose. This approach benefits the subsequent optimization process: the dynamic lip motion yields more natural animation, while the static head pose simplifies landmark-based optimization.

\input{figure_tex/03_03_inference}

\subsection{Optimization} \label{sec:optimization}
From the previous steps, a target video that aligns with both the 3D character and the input audio can be generated. The purpose of optimization is to estimate the blendshape parameters, $B \in \mathbb{R}^{F \times \mathbf{N}}$, that correctly reflect the motion of generated 2D video for final 3D speech animation. Here, $F$ is the number of frames and $N$ is the number of blendshape parameters. The objective functions for the optimization consist of talk landmark loss, asymmetric mouth opening loss, and regularization loss. After the optimization, a Gaussian filter is applied to further improve the naturalness of speech animation by reducing temporal jittering. Additionally, to further enhance the realism of the speech animation, we can optionally add random eye blinks or facial expressions, as our method is based on blendshapes. The full optimization process is summarized in Figure~\ref{fig:optimize}.

\input{figure_tex/03_04_optimize}

The talk landmark loss ensures that the talk-related landmarks of the 3D character align accurately with those in the generated 2D video. To automatically obtain the landmarks of the 3D character, we first render the neutral 3D character into an image $I_0$. From this image, we extract 2D talk-related landmarks using a pre-trained landmark estimator~\cite{yang2025x} denoted by $\phi$. For each 2D talk-related landmark, we identify the corresponding vertex on the character by casting a ray from each 2D landmark and selecting the nearest vertex to the intersection point. \fixq{This process is performed only once with the neutral mesh, thus frame-by-frame instability is avoided. It should be noted, however, that large occluding components such as masks or highly stylized dental geometry may interfere with ray-casting. In such cases, these components are temporarily removed prior to vertex selection, or manual vertex assignment is performed.} This gives the indices of the talk-related landmark vertices. Next, we compute the 2D locations of the talk-related landmark vertices for the deformed character created by blendshape parameters, $B_f \in \mathbb{R}^{\mathbf{N}}$, at frame $f$ (where $1 \leq f \leq F$), using blendshape matrix $M_b$ and projection matrix $P$. This can be denoted as $P\left[ \bigl(M_b \cdot B_f\bigr)_{\text{talk}}\right]$. The talk landmark loss is then defined as the mean-squared error between these 2D talk landmark positions of the 3D character and the talk landmark positions in the f'th frame of the generated video $\hat{I}_{f}$.

Natural human motion often exhibits slight head movements such as small nods or shakes during speech. These movements can cause minor shifts in facial alignment, which need to be accounted for to maintain accurate landmark matching. Therefore, we compute this shift between $\hat{I}_{f}$ and ${I}_{0}$ using an image homography~\cite{raguram2012usac} derived from the estimated landmark positions, and apply the resulting transformation matrix $H$ to the estimated landmarks in the videos before calculating the talk landmark loss. When computing the homography using landmarks, we include only those that do not vary with facial expressions. Thus, we filter out the expression-variant landmarks before obtaining $H$. A detailed explanation of this filtering is provided in Section~\ref{sec:filter}. The objective for the talk landmark loss can be expressed as follows:
\begin{equation} ~\label{eq:talkloss}
L_{talk} = \lVert P\left[ ({{M_b}\cdot{B_f})_{talk})} \right]-H(\phi(\hat{I}_{f})_{talk}) \rVert_{2}^{2}.
\end{equation}

\renewcommand{\arraystretch}{1.35}
\begin{table*}[t]
    \centering
    \caption{Character configurations for experiments. Each character has different number of vertices, number of meshes, number of blendshape parameters, mesh type, and style.}
    \label{tab:characters}
    \setlength{\tabcolsep}{3pt}
    \scalebox{1}{
        \begin{tabular}{cccccc}
            \hline
            \textbf{Character} & Morphy & Malcolm & Victor & Emily & VMan \\ 
            & 
            \includegraphics[width=1.8cm]{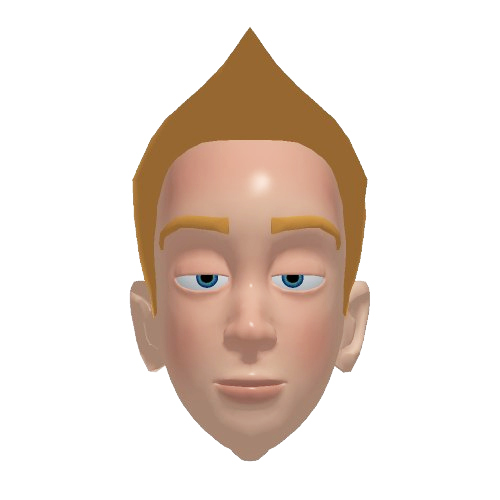} &
            \includegraphics[width=1.8cm]{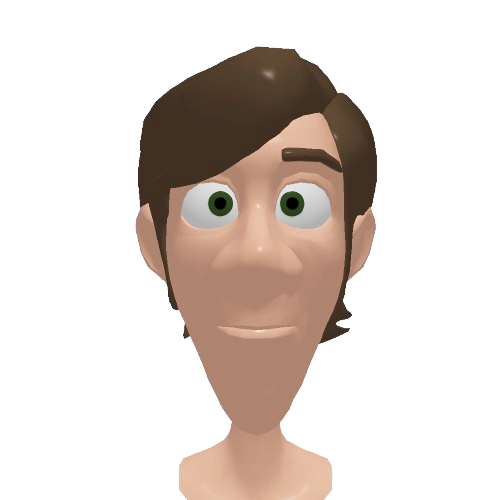} &
            \includegraphics[width=1.8cm]{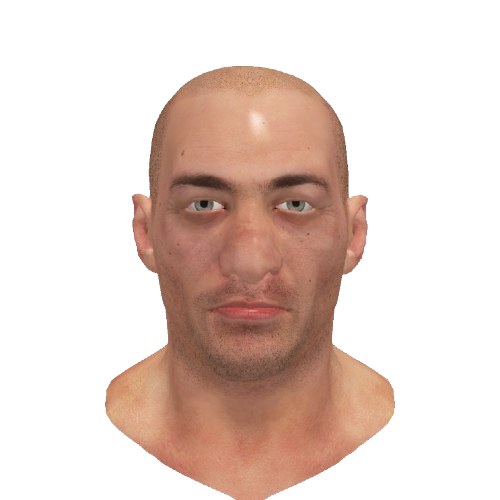} &
            \includegraphics[width=1.8cm]{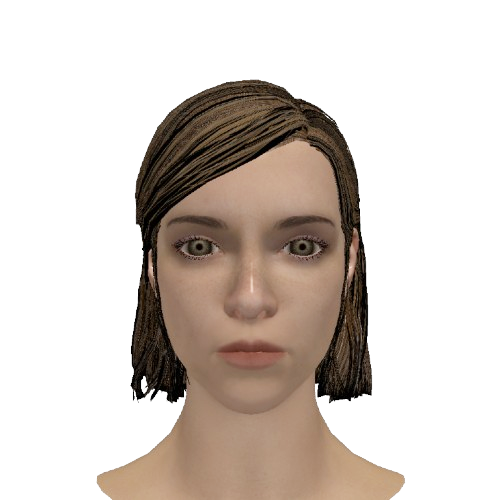} &
            \includegraphics[width=1.8cm]{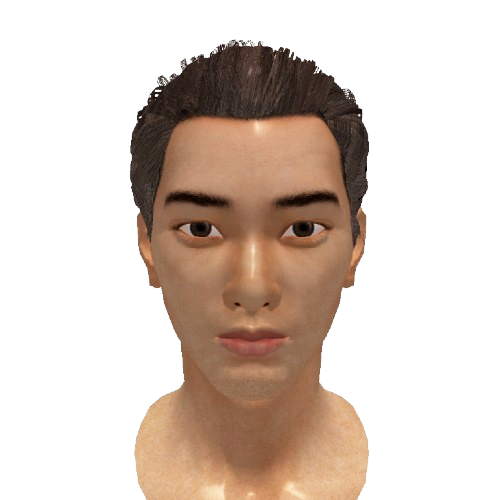} \\ \hline
            \textbf{\# Vertex}     & 4,862 & 4,542     & 20,104  & 33,966 & 241,981 \\ \hline
            \textbf{\# Mesh}       & 7     & 4     & 1       & 5      & 5       \\ \hline
            \textbf{\# Param}      & 46    & 32    & 45      & 121    & 101     \\ \hline
            \textbf{Mesh Type}  & \rule[-2.5ex]{0pt}{6.5ex} \makecell{Triangular/Quad} & \makecell{Triangular/\\Quad/N-Gon} & Triangular & Triangular & \makecell{Triangular/Quad} \\ \hline
            \textbf{Stylized}   & Yes   & Yes   & No      & No     & No      \\ \hline
        \end{tabular}\vspace{-2mm}
    }
\end{table*}

One particular challenge faced by many speech-driven facial animation methods is the lack of high-frequency motion~\cite{xing2023codetalker}. To address this, we adopt a new asymmetric mouth opening loss function that enforces alignment between the 3D character's mouth opening and that of the subject in the video. Initially, we measure the distance between the center landmark among all upper-lip landmarks  $(M_b\cdot B_f)_{up}$ and the center landmark among all lower-lip landmarks  $(M_b\cdot B_f)_{low}$ and adjust it to be similar to the corresponding distance observed in the video. This simple step ensures that the 3D character’s range of mouth opening matches with that captured by the video. On top of this, to avoid smaller opening of the mouth than that of the video, we apply a weight $w_{asym}\;(w_{asym}>1)$. This can be expressed as follows:
\begin{equation}~\label{eq:open1}
\begin{aligned}
L_{open} &= 
\begin{cases}
\|\text{MO}\|_2^2, & \text{if } \text{MO} > 0,\\[4pt]
w_{asym} \cdot \|\text{MO}\|_2^2, & \text{otherwise},
\end{cases} \\[7pt]
\text{where } \text{MO} &= P\Bigl[(M_b\cdot B_f)_{up} - (M_b\cdot B_f)_{low}\Bigr] \\
&\quad - H \Bigl(\phi(\hat{I}_{f})_{up} - \phi(\hat{I}_{f})_{low}\Bigr).
\end{aligned}
\end{equation}
Finally, regularization loss is applied to ensure that blendshapes unrelated to speech, such as those for eyebrows or ears, are not altered. The regularization can be expressed as follows:
\begin{equation}~\label{eq:reg}
L_{reg} = \lVert B_f \rVert_{1}.
\end{equation} 
The optimization process minimizes the loss function $L_{optim}$ to obtain the optimal blendshape parameters $B^*_f$. This can be written as follows:
\begin{align}~\label{eq:optimize}
L_{optim} &= \lambda_{talk}L_{talk} + \lambda_{open}L_{open} + \lambda_{reg}L_{reg},\\[6pt]
B^*_f        &= \arg\min_{B_f}\bigl(L_{optim}(M_b \cdot B_f)\bigr).
\end{align}
Here, each $\lambda$ is a weighting factor that balances its corresponding loss term.

%% file: figure_tex/03_00_baseline.tex
\begin{figure}
  \includegraphics[width=\linewidth]{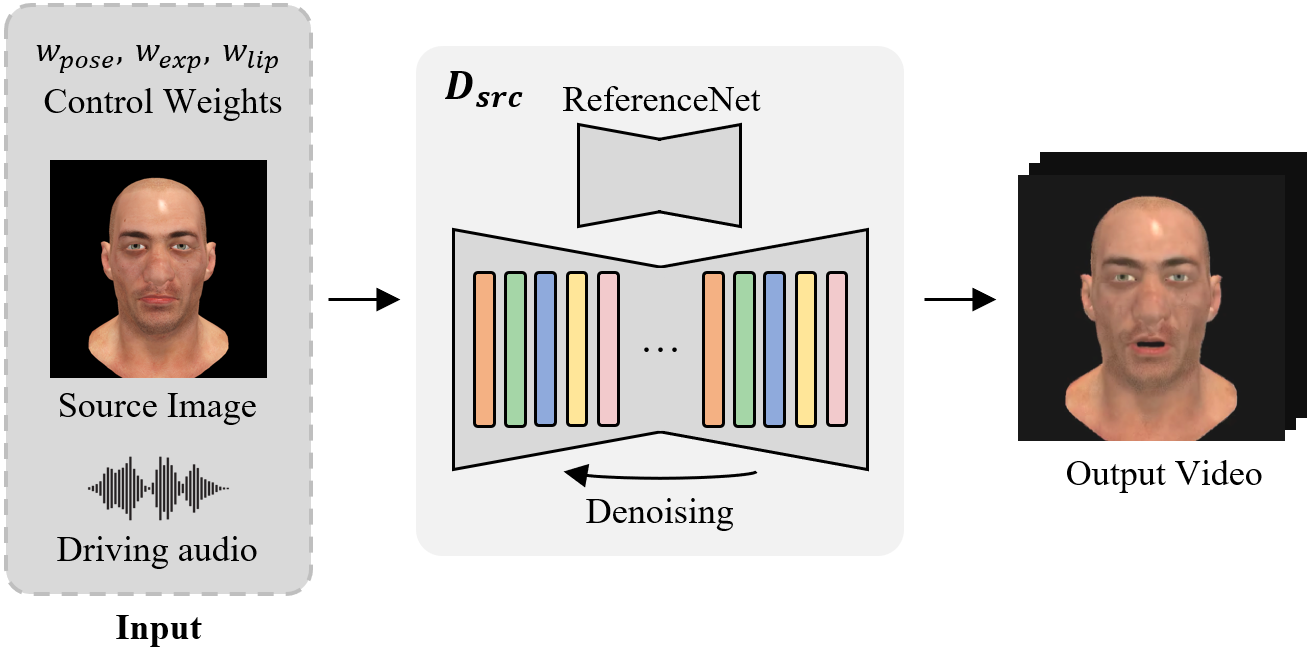}
  \caption{Overview of baseline audio-driven video diffusion model $D_{src}$. Because $D_{src}$ adopts $w_{pose}$, $w_{exp}$, and $w_{lip}$ to control the degree of motion, AnyTalk is capable of generating  dynamic lip movements while maintaining minimal head rotation.}
  \label{fig:base_hallo}
\end{figure}

%% file: figure_tex/03_01_domain_gap.tex
\begin{figure}
  \includegraphics[width=\linewidth]{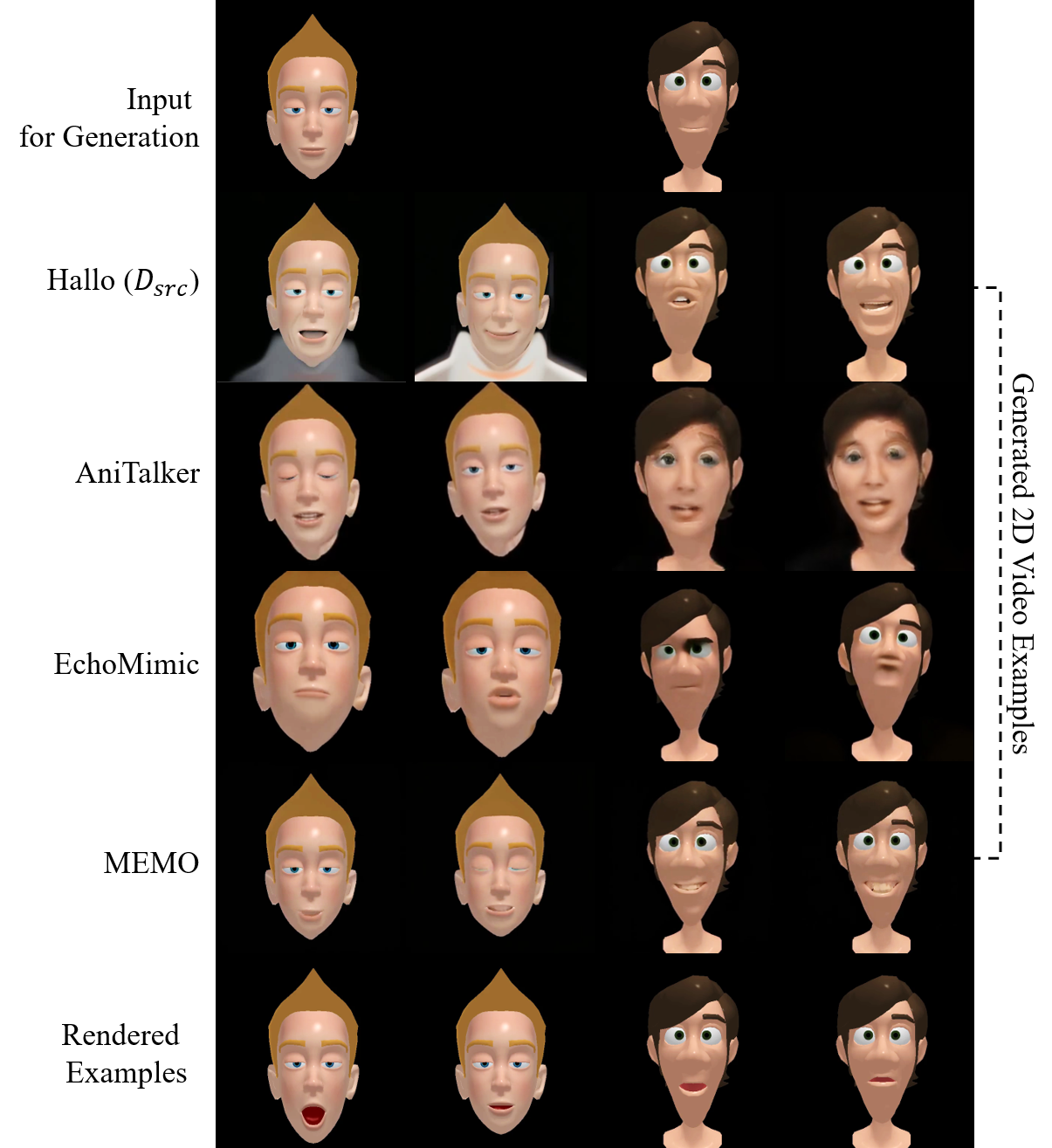}
  \hspace{-2mm}
  \caption{Domain gap between generated and rendered images. From an input image (left), a video is generated using existing talking-head generation models~\cite{xu2024hallo,zheng2024memo,liu2024anitalker,chen2025echomimic} (middle). A clear mismatch can be observed between their appearances and  the rendered images (right).}
  \label{fig:domain_gap}
\end{figure}

%% file: figure_tex/03_02_training.tex
\begin{figure*}[t]
  \includegraphics[width=\linewidth]{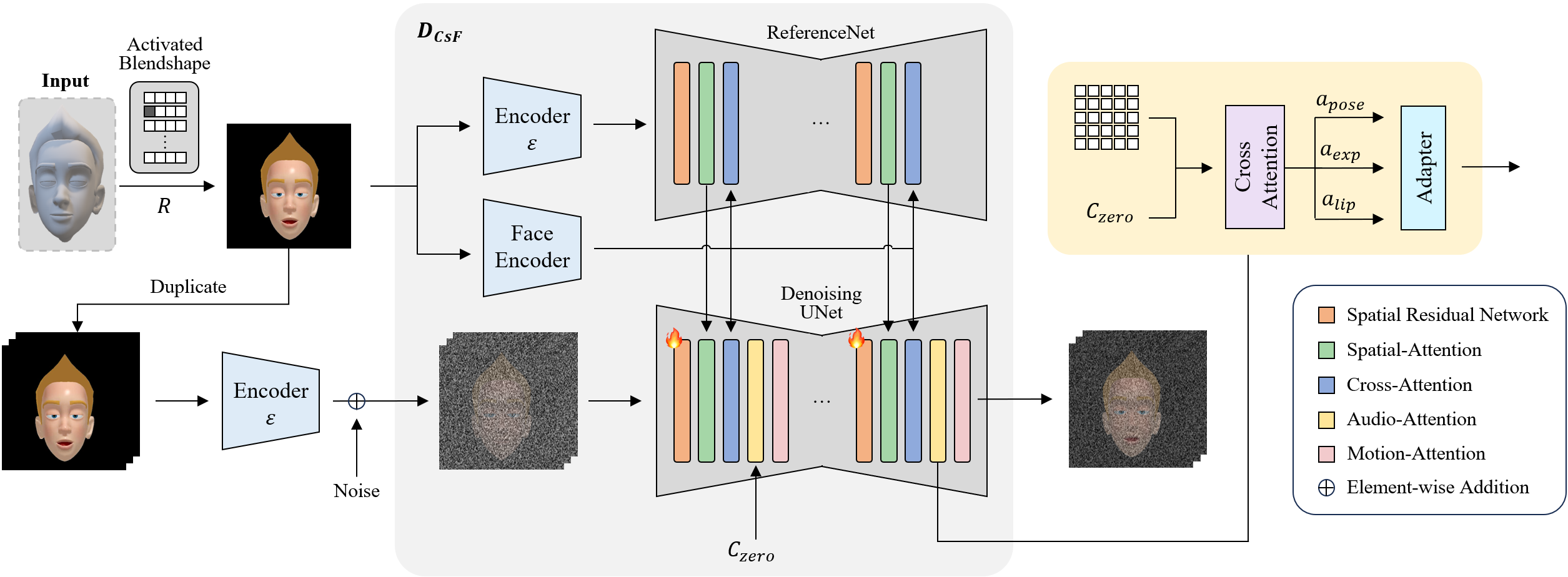}
  \caption{Finetuning process of $D_{CsF}$. The rendered images of the target 3D character and the zero condition $C_{zero}$ are used to form a pair for fine-tuning the model. To preserve the motion prior, only the spatial residual network of denoising UNet is trained (flame icon), while the other layers remain frozen.}
  \label{fig:training}
\end{figure*}

%% file: figure_tex/03_03_inference.tex
\begin{figure}[t]
\centering
  \includegraphics[width=\linewidth]{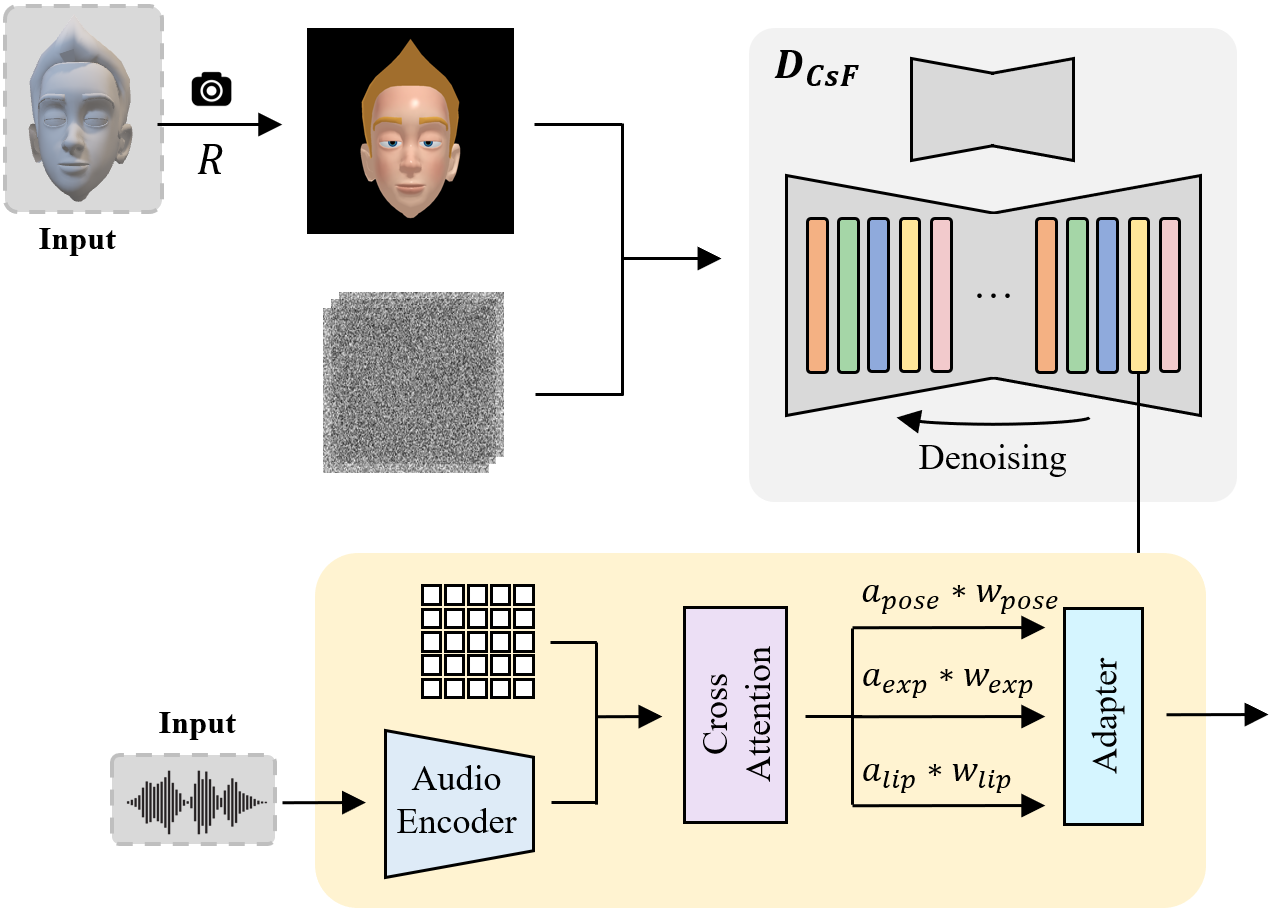}
  \caption{
  Inference of $D_{CsF}$  . Non-zero speech passes through the audio encoder to provide motion information, while the latent representations $a_{pose}$, $a_{exp}$, and $a_{lip}$ are scaled by the corresponding factors $w_{pose}$, $w_{exp}$, and $w_{lip}$ to achieve dynamic lip motion and a stable head pose.}\label{fig:inference}
\end{figure}

%% file: figure_tex/03_04_optimize.tex
\begin{figure}[t]
\centering
  \includegraphics[width=\linewidth]{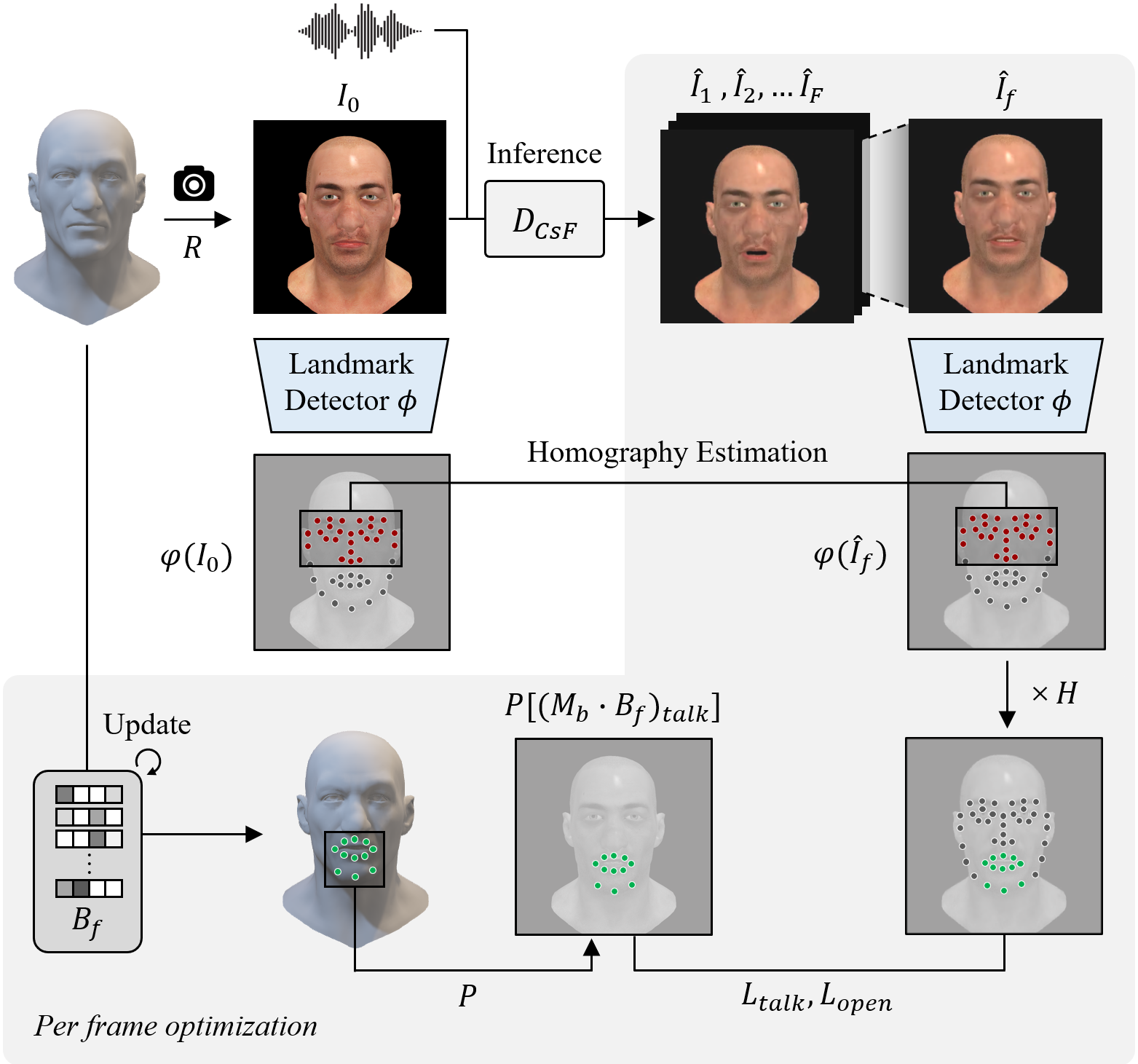}
  \caption{Optimization process. Blendshape parameters are optimized so that the target 3D character accurately follows the generated video when rendered. }
  \label{fig:optimize}
\end{figure}

%% file: sec/4_experiments.tex
\section{Experiments} 
\subsection{Implementation Detail}\label{sec:implementation}
We implemented AnyTalk and conducted all training and inference on a computer with a single Nvidia A6000 GPU. For $CsF$, images were rendered in a frontal view with each blendshape activated. When a mouth related blendshape was activated, the images were duplicated four times. The learning rate was set to 1e-6, and the total training time was 20 minutes. We used 14 talk-related landmarks from the lips and chin for $L_{talk}$ in Equation~\eqref{eq:talkloss}. The asymmetric weight $w_{asym}$ used in Equation~\eqref{eq:open1} was set to 3. The weights $\lambda_{talk}$, $\lambda_{open}$, and $\lambda_{reg}$ used in Equation~\eqref{eq:optimize} for optimization were set to 100,000, 8,000, and 10, respectively. Optimization was conducted for 200 iterations, with learning rate of 5e-3.

\subsection{Dataset}
The only requirement of our method is to provide a target character. To conduct experiments, we gathered five different target characters: Morphy(\copyright joshburton.com), Malcolm(\copyright Animschool), Victor (\copyright Faceware Technologies, Inc.), Emily, and VMan. Each character has a different configuration in terms of the number of vertices, the number of constituent meshes, blendshape parameters, mesh type, and style as shown in Table~\ref{tab:characters}. \fixq{Notably, our test suite covers a broad spectrum of visual styles, ranging from highly stylized characters like Malcolm to photorealistic human avatars such as Victor and Emily.} To generate speech animation, we randomly sampled audio clips from the LibriSpeech~\cite{panayotov2015librispeech} dataset and used them for each of the five characters. 

\input{figure_tex/04_01_lmk}

\begin{figure*}[h]
\centering \vspace{-3mm}
  \includegraphics[width=0.95\linewidth]{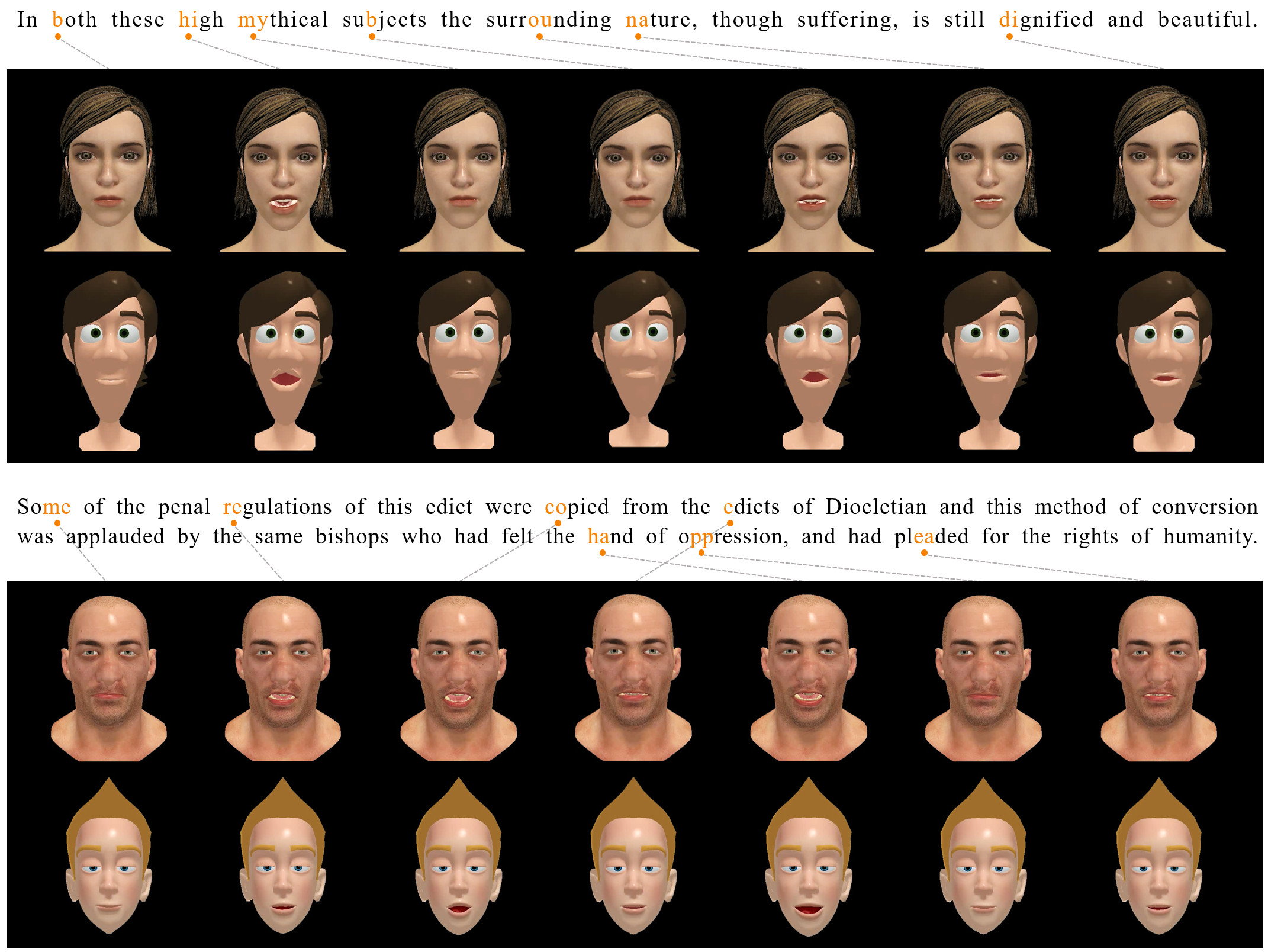}
  \caption{Qualitative results of AnyTalk using two different audio sources. The top two rows shows results of Emily and Malcolm, while bottom two rows shows result of Victor and Morphy.}
  \label{fig:our_results}
\end{figure*}

\begin{figure}[t!]
\centering
\includegraphics[width=0.95\linewidth]{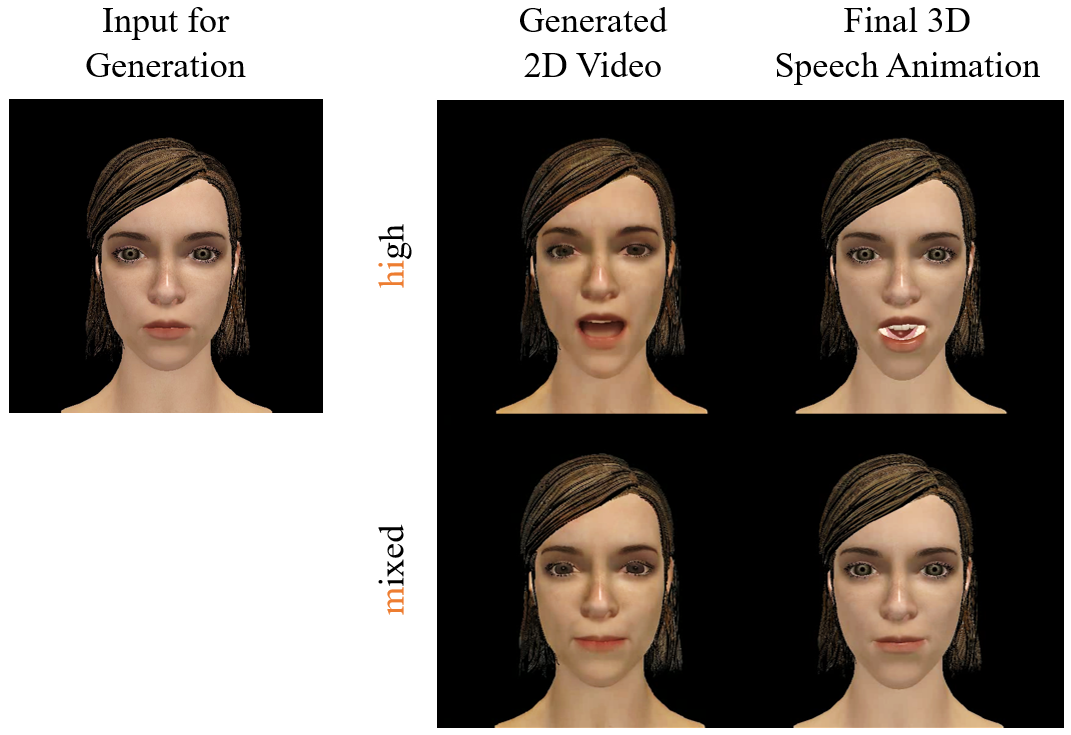}\vspace{-1mm}
  \caption{From input image (left) and audio, 2D video is generated using $D_{CsF}$ (middle). From this video, final 3D speech animation is optimized (right).}
  \label{fig:two_step}
  \vspace{-1mm}
\end{figure}

\subsection{Landmark Filtering} \label{sec:filter}
We conducted an experiment to select the expression-invariant landmarks to be used for estimating the homography between the generated 2D video and the rendered images of the 3D character. In this experiment, we used all five characters and activated each blendshape to generate faces with different expressions. For each landmark, we calculated the average displacement across the various blendshape parameters. This analysis, as shown in Figure~\ref{fig:lmk}, allowed us to distinguish between stable landmarks and those prone to variation. We observed that the landmarks corresponding to the lower eyes, nose, and sides of the head exhibited relatively small shifts, whereas the landmarks on the lips and chin showed the most significant changes. Based on these findings, we decided to filter out the top half of expression-variant landmarks during homography estimation to ensure robust and accurate landmark matching in subsequent processing. Although filtering additional landmarks resulted in a lower average variance, using too few landmarks for homography estimation compromised robustness.

\subsection{Results} \label{sec:results}
We demonstrate our 3D speech animation results on arbitrary characters driven by a provided audio track, spanning a wide range of mesh structures and artistic styles. To evaluate lip synchronization quality, we present seven key frames rendered from a facial animation that aligned with specific phonemes in Figure~\ref{fig:our_results}. For visualization, each phoneme or letter and its corresponding frame are connected with a dotted line. The results indicate that the generated lip movements closely follow the spoken phonemes.

We also visualize both the 2D video generated by $D_{CsF}$ and the optimized final animation results. As shown in Figure~\ref{fig:two_step}, due to the use of $CsF$ that personalize video generation model, the generated video accurately followed the target character driven by the input audio, without visual mismatches. Furthermore, the alignment and optimization process ensures that the final lip motion precisely matches the generated video corresponding to the phonemes of the source audio.


\begin{figure*}[h]
\hspace{-5mm}
\vspace{-3mm}
  \includegraphics[width=0.95\linewidth]{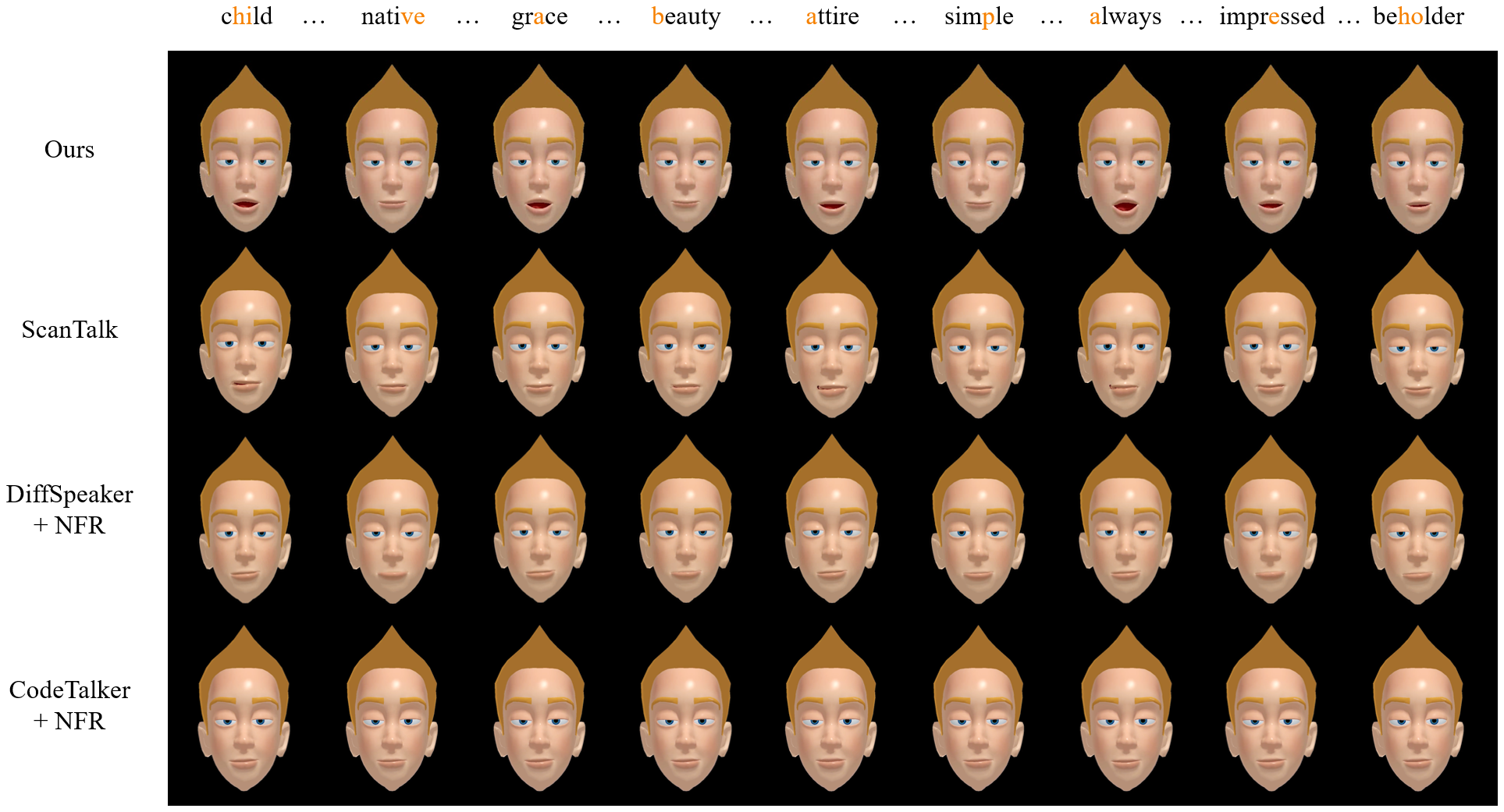}
  \centering
  \caption{\fixq{Comparison with baselines on Morphy. Each phoneme and its corresponding frame are presented for comparison. Ours best follows the given audio with wider mouth opening and precise lip closure, while ScanTalk, DiffSpeaker + NFR, and CodeTalker + NFR exhibit only subtle movement.}}
  \label{fig:baseline1}
\end{figure*}

\subsection{Baseline Comparison} \label{sec:baseline}
\fixq{We compared our approach with ScanTalk~\cite{nocentini2025scantalk}, DiffSpeaker~\cite{ma2024diffspeaker}, and CodeTalker~\cite{xing2023codetalker},} using all 5 characters and 30 audio clips, resulting in a total of 150 speech animations per method. ScanTalk is our closest competitor because it is the only method capable of generating 3D speech animations for diverse mesh structures. DiffSpeaker is a diffusion-based 3D speech animation method that does not require additional information (such as a style video) during inference. \fixq{In addition, CodeTalker is a 3D speech animation method that uses VQ-VAE latent for autoregressive motion generation. Because DiffSpeaker and CodeTalker} can only work with a specific mesh structure on which it was trained, we first generated outputs using the VOCASET mesh structure and subsequently retargeted them with NFR~\cite{qin2023neural}, a state-of-the-art neural retargeting method. Both ScanTalk and NFR require the mesh to be scaled, aligned, and stripped of unnecessary components (e.g., accessories or hair) prior to inference, accordingly, we followed their guidelines for inference. For comparison, we rescaled, realigned, and restored any removed parts to render the meshes in their original form. \fixq{Additional comparison results with an image-based baseline~\cite{choi2024deep} that could not generate motion are presented in Section 2.3 of the supplementary material. It is important to note that our method is personalized to each character, whereas the baseline methods are not. Thus, we also evaluated our approach in a non-personalized setting and reported the results in Section 2.4.2 of the supplementary material.}

\begin{table}[t]
    \centering
    \caption{Quantitative results comparing our method with baselines. Best denoted in bold.}
    \label{tab:quant_baseline}
    \scalebox{1}{
        \begin{tabular}{lccc}
            \hline
            \textbf{Methods} & LSE-D$\downarrow$ & LSE-C$\uparrow$  \\ \hline
            Ours                & \textbf{11.304}    & \textbf{3.155} \\ \hline
            ScanTalk            & 12.152    & 2.395 \\ \hline
            DiffSpeaker + NFR   & 13.857    & 0.665 \\ \hline
            \fixq{CodeTalker + NFR}   & \fixq{13.840}    & \fixq{0.668} \\ \hline
        \end{tabular}
    }
\end{table}

The results of the qualitative comparison are presented in Figure~\ref{fig:baseline1}. The rendering settings were the same for all methods. To evaluate lip synchronization, we present nine specific frames from the synthesized facial animations, similar to those in Section~\ref{sec:results}. As shown in the figure, the lip movements created by our method more precisely follow the given sentence with dynamic motion. \fixq{In contrast, ScanTalk, DiffSpeaker + NFR, and CodeTalker + NFR exhibited only subtle movements that did not faithfully follow the intended phonemes.} This discrepancy may be attributed to differences in the characters on which ScanTalk and NFR were trained. In contrast, our method successfully handled meshes with varying numbers of vertices or stylized characters, without requiring any additional processing. 

Because no ground-truth pairs of audio and 3D animations are available for quantitative comparison, we evaluated lip synchronization using the Lip Sync Error Distance (LSE-D) and Lip Sync Error Confidence (LSE-C) metrics \cite{prajwal2020wav2lip}, which do not require ground-truth data. These metrics can be expressed as follows:
\begin{align*}
d_n(k) &= \bigl\lVert v_n - a_{n+k} \bigr\rVert_2, \quad \mathrm{LSE\text{-}D} = \frac{1}{N}\sum_{n=1}^N \min_{k} \,d_n(k), \\[4pt]
\mathrm{LSE\text{-}C} &= \frac{1}{N}\sum_{n=1}^N \Bigl(\,\mathrm{median}_{k}\,d_n(k)\;-\;\min_{k}\,d_n(k)\Bigr).
\end{align*}
Here, $v_n$ is the SyncNet~\cite{chung2017out} video embedding for the n-th clip, $a_{n+k}$ is the SyncNet audio embedding, and $d_n(k)$ is their \fixq{Euclidean} distance. As shown in Table~\ref{tab:quant_baseline}, \fixq{ours} achieved the best results on both metrics, while ScanTalk achieved the second best performance. \fixq{DiffSpeaker + NRF and CodeTalker + NFR produced very similar results both qualitatively and quantitatively, likely due to the use of same retargeting.}


\input{figure_tex/04_02_vid_ablation}
\input{figure_tex/04_03_ablation}

\subsection{Ablation Study}
We conducted a series of ablation studies to validate the proposed video generation and optimization methods by modifying each component. We used the same audio as in Section~\ref{sec:baseline} and employed two characters, VMan and Malcolm. Our Ablation studies mainly consists of two parts. Firstly, ablating fine-tuning process for personalized video generation, secondly, ablating optimization process for final speech animation. 

On the video generation process, we compared \fixq{ours} with three variations. First, we ablated $CsF$, while retaining $w_{pose}$, $w_{exp}$, and $w_{mouth}$, as described in Section~\ref{sec:inference}. Second, we ablated $C_{zero}$, and instead used random audio to match with the duplicated images of the character for $CsF$. Third, we compared \fixq{ours} with an alternative that did not freeze the spatial, temporal, and cross attention modules when fine-tuning. For the optimization process, we compared with another three variations. First, we ablated the loss term $L_{talk}$. Second, we evaluated landmark filtering for homography estimation by comparing \fixq{ours with} an approach that uses all 68 facial landmarks without filtering. Third, we tested an additional filtering strategy that uses only the 10 landmarks with the lowest average distance.

\subsubsection{Ablation Study on Video Generation}
First, we compared variants of the 2D video generation process. Because video generation is the first stage of our method, accurately producing a talking‐head video that does not deviate from the target character while following input audio is crucial for achieving the final speech animation. As shown in Figure \ref{fig:vid_ablation}, \fixq{ours} generates the lip shapes correctly and without artifacts. In contrast, the variant without $CsF$ produced inconsistent appearances, such as squeezed pupils and double chins. The variant without $C_{\text{zero}}$ yielded less expressive lip motion and failed to open the mouth properly because its training did not correctly disentangle the no‐motion signal from the speech signal. Finally, the variant without freezing modules exhibits the poorest performance: it produced minimal motion with no mouth opening, likely due to overfitting of the attention modules.

\subsubsection{Ablation Study on Final 3D Speech Animation}
The ablation study results for 3D speech animation are presented in Figure~\ref{fig:ablation}. In the variant without $CsF$, lip movements did not match the given audio due to mismatches during video generation, whereas the variants without $C_{zero}$ and without freezing modules failed to open the mouth sufficiently, owing to the entanglement of motion signals and overfitting in the motion module. The variant without $L_{talk}$ exhibited artifacts on the chin because blendshapes were not correctly estimated in the absence of talk-related landmark optimization, relying solely on a sparse mouth-opening loss. In the filtering-process ablation, the variant without landmark filtering barely opened the mouth for the phoneme “be,” whereas adding the filtering step yielded incorrect blendshapes due to homography estimation errors showing that 10 landmarks are not sufficient to estimate homography matrix robustly. In contrast, \fixq{ours} produced results that closely followed the spoken phonemes without artifacts and correctly opened the mouth. This superiority is also verified by the quantitative results reported in Table~\ref{tab:quant_ablation}. Ours achieved the best scores for LSE-D and the second best score for LSE-C. Without $CsF$ achieved the best score for LSE-C but did not perform well for LSE-D.

\paragraph{Photometric Optimization.} In addition to the ablation study of each proposed component, we evaluated two additional variants to further investigate whether geometric optimization through 3D landmark supervision is indeed the optimal choice for speech-driven facial animation. We defined a photometric loss $L_{photo}$ based on LPIPS~\cite{zhang2018lpips} to capture the perceptual similarity between the rendered character and the generated video. First, we replaced all 3D landmark-based losses ($L_{talk}$ and $L_{open}$) with $L_{photo}$. When solely using $L_{photo}$ to encourage the rendered character to match the generated video, the results exhibited no or only subtle movement. Second, we added $L_{photo}$ to our original loss terms. This combined variant produced results similar to ours, with slight degradation in both quantitative and qualitative metrics, as shown in Table~\ref{tab:quant_ablation} and Figure~\ref{fig:ablation}. Moreover, because incorporating the photometric loss requires rendering the character with a differentiable renderer at every iteration, the animation generation time increased by a factor of 2.96 (9.32 s vs. 3.12 s per frame). These results confirm that photometric loss is a highly ambiguous objective for blendshape optimization, offering no additional performance gain and significantly slowing down the optimization process.

\begin{table}[h]
    \centering
    \caption{Quantitative results from the ablation study. The best and the second best results are denoted in \textbf{bold} and \underline{underlined}.} 
    \label{tab:quant_ablation}
    \setlength{\tabcolsep}{10pt}
    \scalebox{1}{
        \begin{tabular}{lccc}
            \hline
            \textbf{Methods} & LSE-D$\downarrow$ & LSE-C$\uparrow$  \\ \hline
            Ours                & \textbf{10.695}     & \underline{3.397}   \\ \hline
            w/o $CsF$    & 11.207    & \textbf{3.451}   \\ \hline
            w/o $C_{zero}$    & 11.203    & 2.770 \\ \hline
            w/o freezing module    & 14.237    & 1.228 \\ \hline
            w/o $L_{talk}$      & 11.044    & 2.922 \\ \hline
            w/o landmark filtering    & \underline{10.777}    & 3.019 \\ \hline
            w/ additional filtering   & 12.397    & 2.352 \\ \hline
            replaced w/ $L_{photo}$ & 15.154 & 0.293 \\ \hline
            w/ $L_{photo}$ & 10.813 & 3.364 \\ \hline
             
        \end{tabular}
    }
\end{table}

\paragraph{Effect of Fine-tuning Strategies for $CsF$.}
\fixq{To substantiate the design of $CsF$ and evaluate its sensitivity to different fine-tuning configurations, we conduct a controlled study varying which UNet blocks are frozen or trained. Our hypothesis is that fine-tuning purely spatial layers on static images while zeroing the audio embedding and explicitly freezing temporal and audio-attention modules enables high-fidelity appearance transfer without overwriting the pre-trained motion priors. To quantify this balance between motion prior retention and appearance transfer, we compare our proposed setting against three variants: fine-tuning with audio-attention unfrozen, fine-tuning with motion-attention unfrozen, and fine-tuning all UNet layers (w/o freezing). As shown in Table~\ref{tab:quant_ablation2}, exposing the audio attention leads to degraded performance which may be due to overfitting zeroed out audio embedding. When exposing the motion-attention modules, the degradation becomes even more pronounced. This is likely because training temporal layers on static data forces the motion priors to collapse, as the model erroneously learns to minimize frame-to-frame variance. Fine-tuning all layers results in the most significant drop in performance, demonstrating catastrophic forgetting of the audio-driven dynamics. In contrast, by restricting updates strictly to the spatial residual blocks, our strategy successfully injects character-specific appearance features while safely disentangling them from the model's foundational temporal and audio-sync priors.}

\begin{table}[t]
    \centering
     
    \caption{\fixq{Quantitative results for fine-tuning UNet blocks. The best and the second best results are denoted in \textbf{bold} and \underline{underlined}, respectively}.}
    \vspace{-2mm}
    \label{tab:quant_ablation2}
    \setlength{\tabcolsep}{10pt}
    \scalebox{1}{
        \begin{tabular}{lccc}
            \hline
            \textbf{Methods} & LSE-D$\downarrow$ & LSE-C$\uparrow$  \\ \hline
            Ours               & \textbf{10.695}     & \textbf{3.397}   \\ \hline
            w/ audio-attention tuning & \underline{10.796} & \underline{3.307} \\ \hline
            w/ motion-attention tuning & 10.932 & 3.149 \\ \hline
            w/o freezing module    & 14.237    & 1.228 \\ \hline
        \end{tabular}\vspace{-3mm}
    }
\end{table}

\paragraph{Effect of Head Pose Stabilization.}
\fixq{While natural head movements are typically desirable for 2D speech animation, our framework lifts the video into a 3D representation via optimization. In this context, significant head motion is detrimental to the stability and accuracy of the 3D reconstruction. To evaluate this, we conducted an additional ablation study varying the pose weight $w_{pose}$, which controls the degree of stabilization. As shown in Figure~\ref{fig:ablation-wpose}, larger values of $w_{pose}$ (indicating more head motion) lead to a degradation in lip-sync quality, as reflected by the LSE-D and LSE-C. This suggests that the model struggles to accurately decouple fine facial deformations from global rigid transformations when large head motion exists. It is important to note that even when $w_{pose}$ is high, the final 3D-optimized output does not exhibit head motion due to the constraints of the subsequent optimization process. However, enforcing stabilization during the initial generation phase (where $w_{pose}=0$) provides a cleaner source for the 3D uplifting process, ultimately yielding the highest motion fidelity.}

\input{figure_tex/04_04_ablation2}
%

\subsection{User Study}
We conducted a user study to compare the perceptual quality of animations produced by our method against two baselines. Participants answered two questions regarding perceptual naturalness and lip synchronization. \fixq{A total of 21 participants (10 males, 11 females), aged 23 to 35, took part in the web-based study, evaluating our method, ScanTalk~\cite{nocentini2025scantalk}, DiffSpeaker + NFR~\cite{ma2024diffspeaker,qin2023neural}, and CodeTalker + NFR~\cite{xing2023codetalker,qin2023neural}. Each participant was presented with 75 side-by-side video pairs in randomized order using a two-alternative forced-choice format. The results are presented in Table~\ref{tab:user_study}. We applied exact binomial tests against a 50\% chance level to assess the significance of these preferences. For perceptual naturalness, participants chose our method over ScanTalk in 413 of 525 comparisons (78.6\%), a highly significant result with p < 0.001 in the binomial test. For lip synchronization, our animations were preferred in 403 of 525 trials (76.8\%), also with p < 0.001. Comparisons with DiffSpeaker + NFR and CodeTalker + NFR also yielded significant improvements for all metrics with p < 0.001. These results confirm that our approach produces perceptually more natural and better lip-synced speech animations than ScanTalk, DiffSpeaker + NFR, and CodeTalker + NFR.}

\begin{table}[h]
    \centering
    \caption{Selection ratio of our method compared to the baselines.}
    \vspace{-2mm}
    \label{tab:user_study}
     
    \scalebox{1}{
        \begin{tabular}{lcccc}
            \hline
            \textbf{Methods} & Naturalness (\%) & Lip-sync(\%) \\ \hline
            Ours vs. ScanTalk               & 78.6     & 76.8 \\ \hline
            Ours vs. DiffSpeaker + NFR      & 99.6     & 99.4  \\ \hline
            Ours vs. CodeTalker + NFR       & 99.8     & 99.8  \\ \hline
        \end{tabular}
        }
\end{table}

%% file: figure_tex/04_01_lmk.tex
\begin{figure}[t]
\centering
  \includegraphics[width=\linewidth]{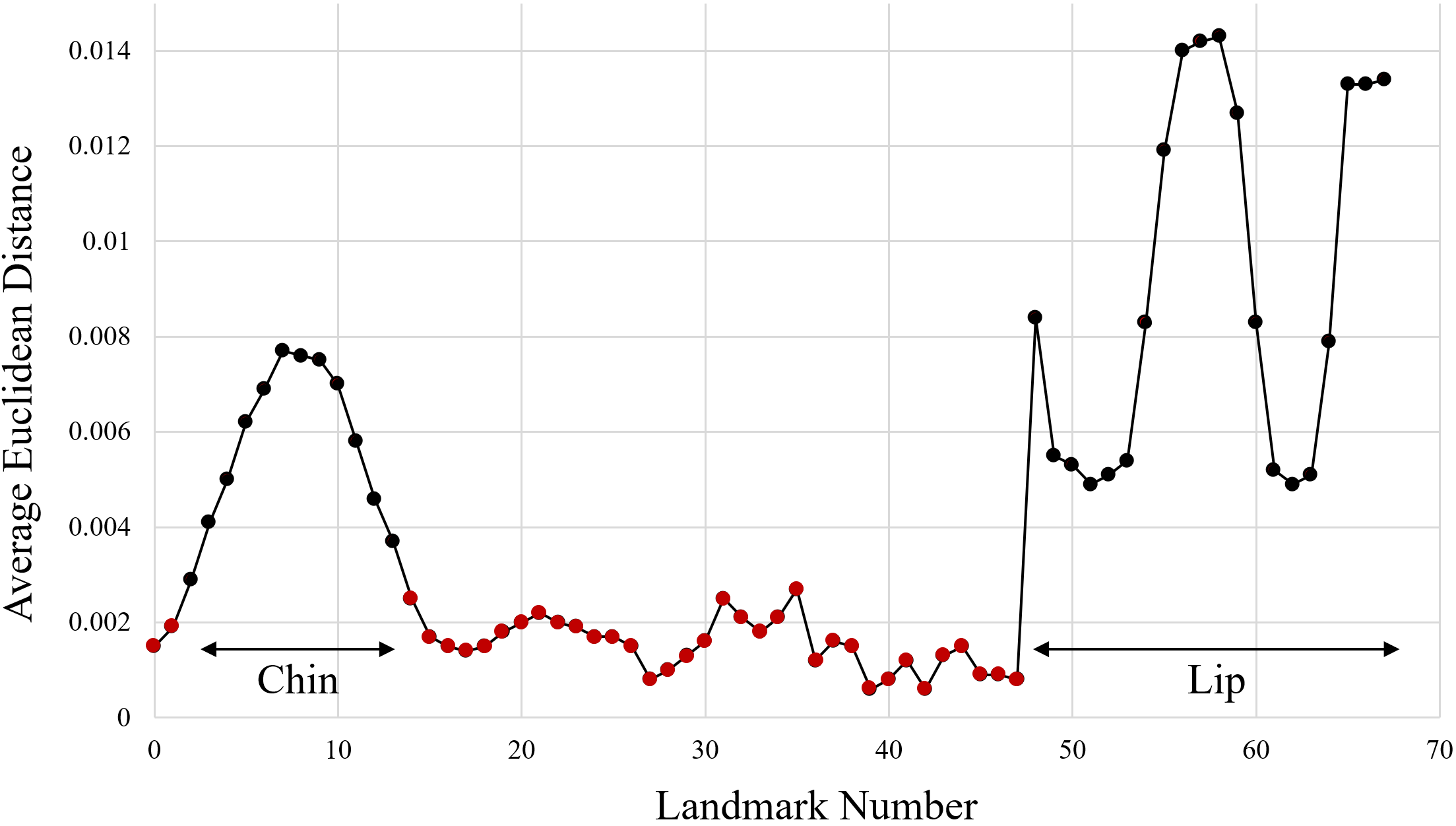}
  \vspace{-2mm}
  \caption{Lip and chin landmarks exhibit the high average displacement, suboptimal for robust homography estimation.}
  \label{fig:lmk}
\end{figure}

%% file: figure_tex/04_02_vid_ablation.tex
\begin{figure}[t!]
\includegraphics[width=1\linewidth]{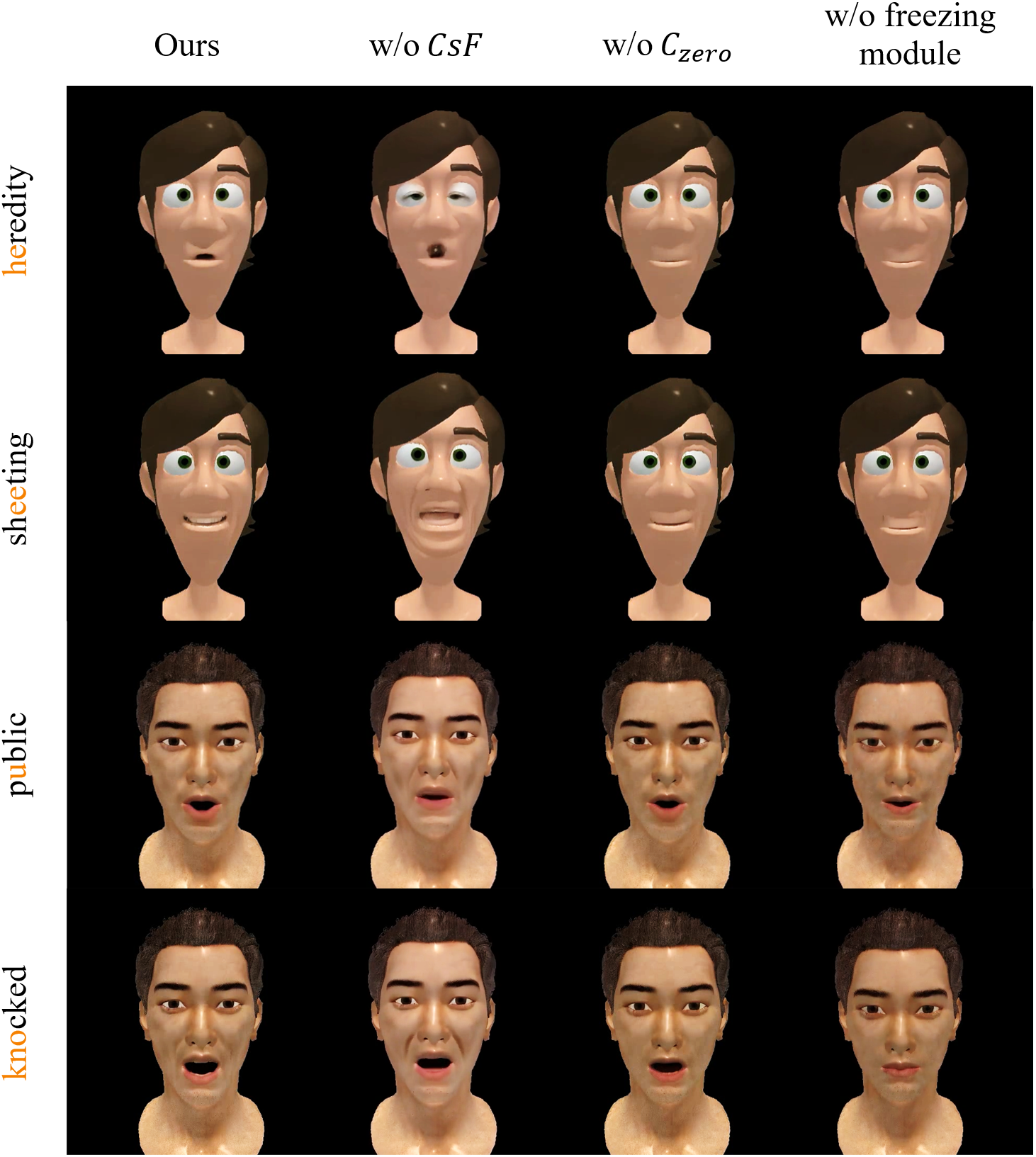} 
\vspace{-2mm}
  \caption{Visual results of generated video. Ours correctly generated the lip shape without artifacts. On the other hand, w/o $CsF$ made double chin, pinched eye, which does not match with the source character. Variants w/o $C_{zero}$ and w/o freezing module produced small and subtle lip motion.}
  \label{fig:vid_ablation}
\end{figure}

%% file: figure_tex/04_03_ablation.tex
\begin{figure*}[ht]
\vspace{-2mm}
\hspace{-4mm}
  \includegraphics[width=1\linewidth]{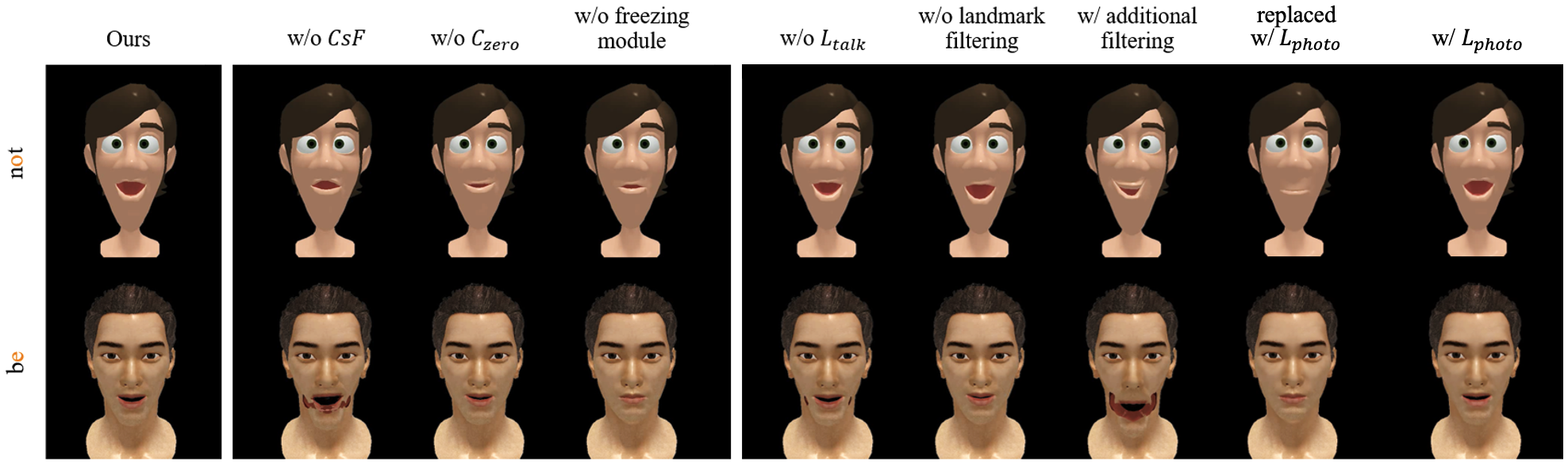}\vspace{-2mm}
  \caption{Results of the ablation study on 3D speech animation. Words corresponding to the rendered images of Malcolm and VMan are presented on the left. While our method correctly produced the lip movements corresponing to the given words, the alternatives produced either only subtle lip movements or artifacts.}\vspace{-2mm}
  \label{fig:ablation}
\end{figure*}

%% file: figure_tex/04_04_ablation2.tex
\begin{figure}[ht]
\vspace{-4mm}
\centering
  \includegraphics[width=0.955\linewidth]{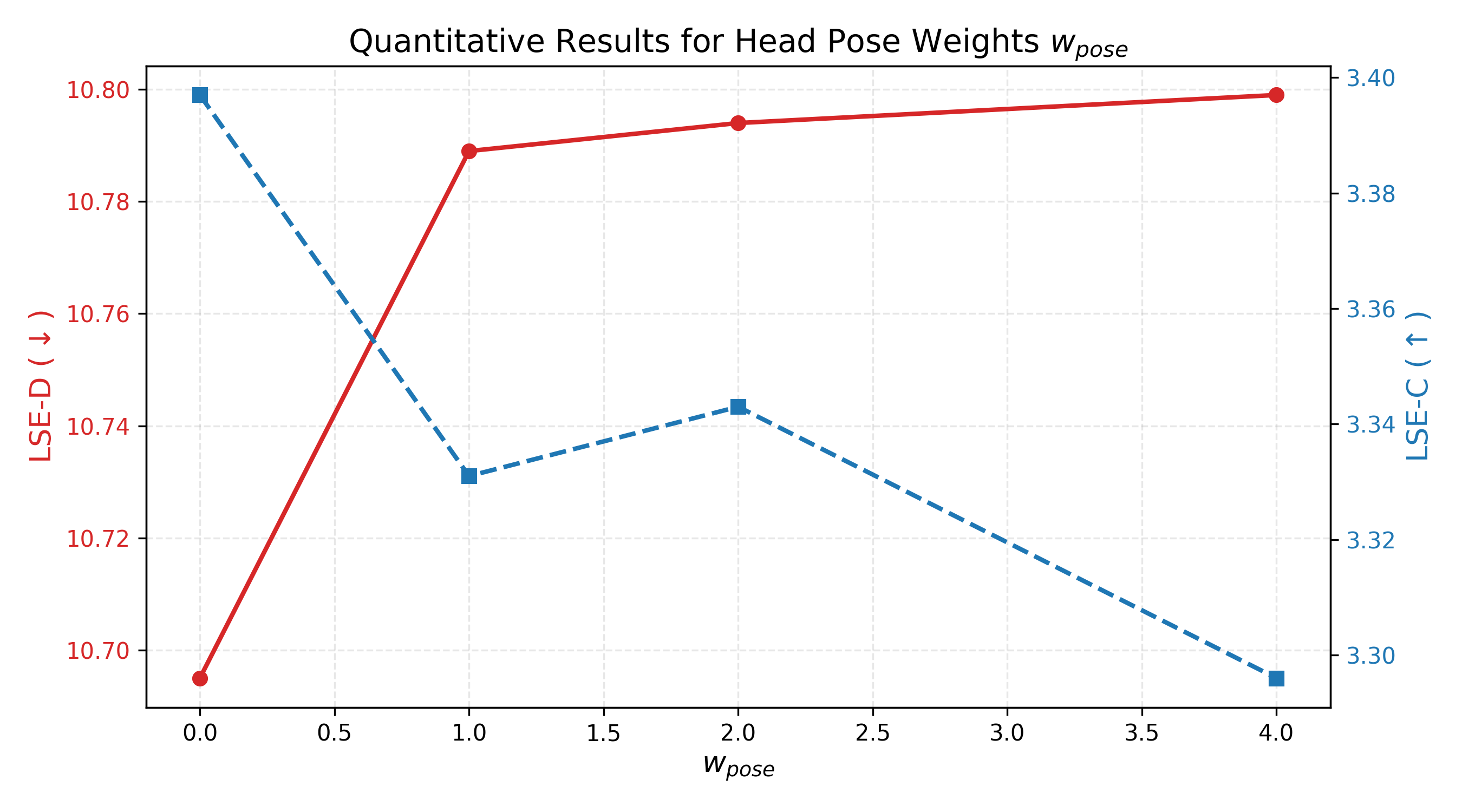}\vspace{-4mm}
  \caption{\fixq{Quantitative results for head pose weights. Lip-sync performance degrades as $w_{pose}$ increase.}}\vspace{-4mm}
  \label{fig:ablation-wpose}
\end{figure}

%% file: sec/5_application.tex
\section{Applications}

\begin{figure}[t!]
    \centering
    \vspace{-2mm}
    \includegraphics[width=0.8\linewidth]{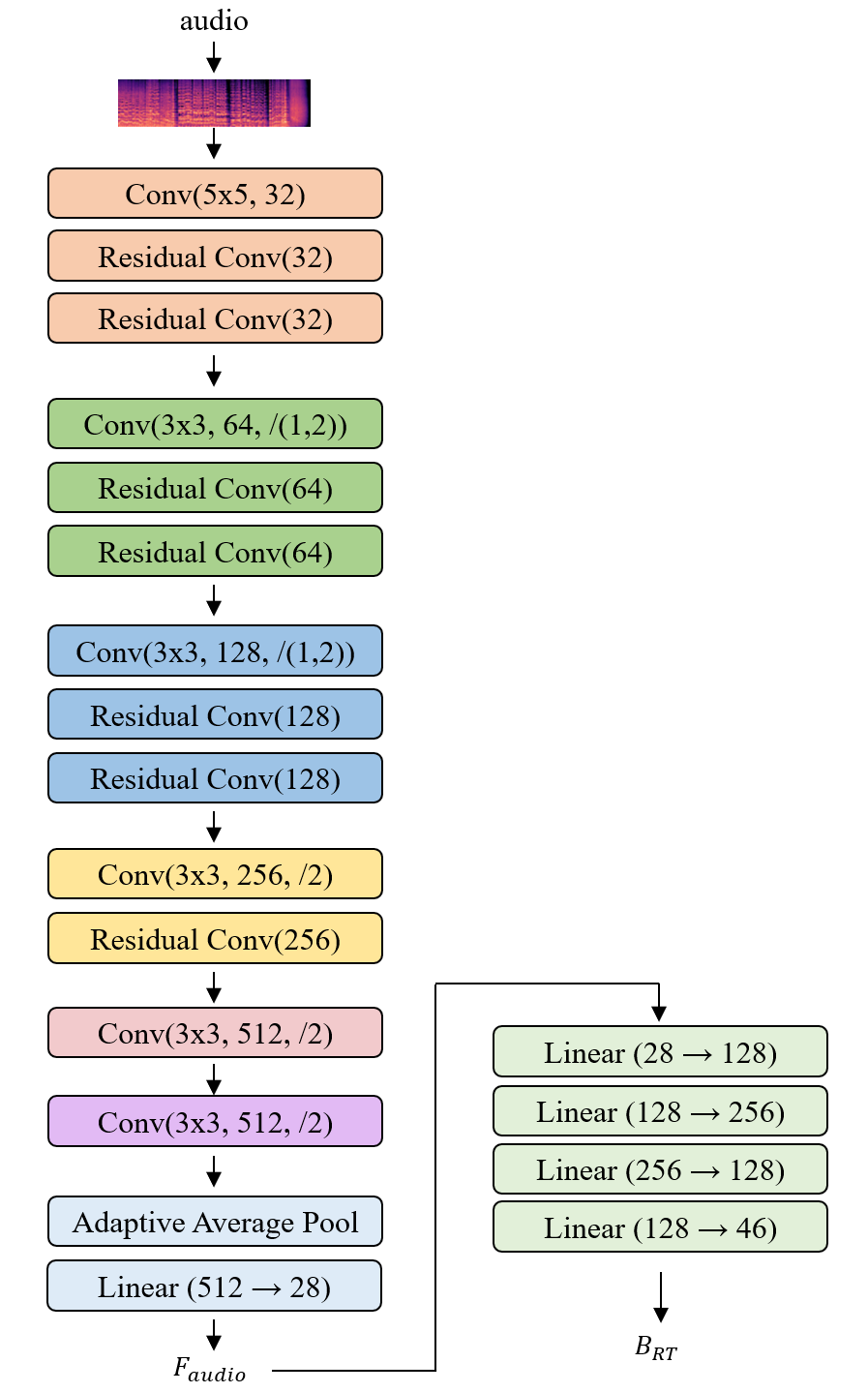}
    \caption{Network architecture for $\text{AnyTalk}_{RT}$.} \vspace{-1mm}
    \label{fig:distil-architecture}
\end{figure}

\subsection{Distillation}
To enable real-time performance for applications, we distilled AnyTalk into a streamlined network, $\text{AnyTalk}_{RT}$. The network architecture is presented in Figure~\ref{fig:distil-architecture}. For training $\text{AnyTalk}_{RT}$, we first generated around 1,600 speech animations using AnyTalk. To effectively distill the optimization based model AnyTalk into the learning based model $\text{AnyTalk}_{RT}$, we applied a feature matching loss $ L_{feat}$ and a blendshape reconstruction loss $L_{recon}$ as follows:
\begin{equation}~\label{eq:open}
\begin{aligned}
L_{distill} &= L_{feat} + \lambda L_{recon}, \quad \text{where} \\
L_{feat} &= F_{audio} - H(\phi (\hat{I}_f)_{talk}) \\
L_{recon} &= B_{RT} - B_f. 
\end{aligned}
\end{equation}
Here, $L_{feat}$ enforces the audio feature of $\text{AnyTalk}_{RT}$ to match with the warped talk-releated landmarks, while $L_{recon}$ enforces the blendshape. The overall distillation process is illustrated in Figure~\ref{fig:distillation}. $\lambda$ is a weighting factor that was set to 400. Training was conducted for 360 epochs using the adamW optimizer ~\cite{loshchilov2017decoupled}, with a batch size of 128, and the learning rate was increased up to 0.008 and decreased using OneCyleLR~\cite{smith2019super}.

\begin{figure}[t]
\centering
  \includegraphics[width=0.9\linewidth]{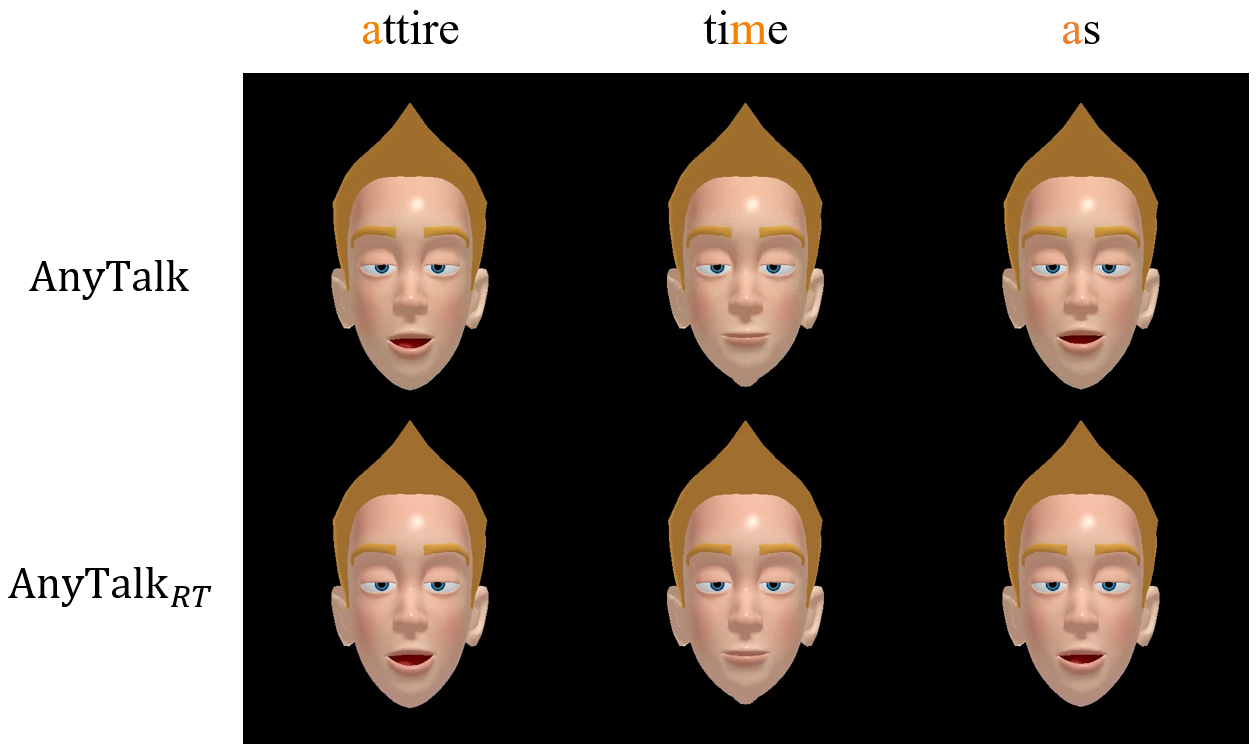}
  \vspace{-3mm}
  \caption{Results of $\text{AnyTalk}$ and $\text{AnyTalk}_{RT}$, correctly following the phoneme of the given audio and depicting similar speech motion.}\vspace{-2mm}
  \label{fig:distill_comparison}
\end{figure}

With the distilled $\text{AnyTalk}_{RT}$, an arbitrary character can be animated in real-time given audio as shown in the Figure~\ref{fig:distill_comparison}. We measured the inference time and lip-sync quality of $\text{AnyTalk}_{RT}$ using Morphy and report the results in Table~\ref{tab:quant-distillation}. Due to the simplified inference, which excludes the optimization phase, and adopts the streamlined network, $\text{AnyTalk}_{RT}$ achieved 9.09 ms per frame (110 frames per second) in a full-precision (32-bit) PyTorch setting. 

\input{figure_tex/05_01_distillation}
\begin{table}[t]
    \centering
    \caption{Quantitative results for $\text{AnyTalk}_{RT}$ compared to $\text{AnyTalk}$. Although distillation enabled real-time performance, the lip-sync metric degraded.} 
    \label{tab:quant-distillation}
    \scalebox{0.95}{
        \begin{tabular}{lcccc}
            \hline
            \textbf{Methods} & Time (per frame) & LSE-D$\downarrow$ & LSE-C$\uparrow$  \\ \hline
            $\text{AnyTalk}_{RT}$ & \textbf{9.09 ms}        & 12.19                 & 2.96   \\ \hline
            $\text{AnyTalk}$ & 3.12 s       & \textbf{11.74}                 & \textbf{3.24}   \\ \hline \vspace{-3mm}
        \end{tabular}
    }
\end{table}

\begin{figure}[t]
  \centering
  \includegraphics[width=0.95\linewidth]{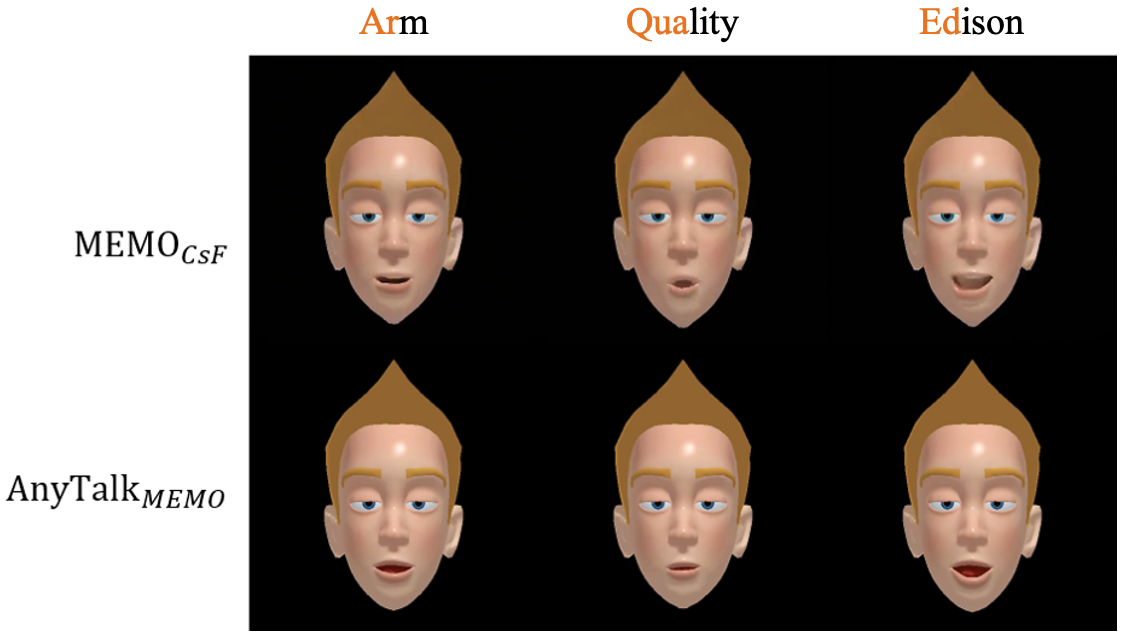}
  \caption{\fixq{Qualitative results of MEMO\textsubscript{CsF} and AnyTalk\textsubscript{\textit{MEMO}}, showing the generalability of AnyTalk Framework.}}
  \label{fig:memo_extension}
\end{figure}

\subsection{Extension to Different Video Generation Model}
\label{sec:extension_baseline}

\fixq{To demonstrate the generality of our $CsF$ and optimization pipeline, we applied the same procedure to MEMO~\cite{zheng2024memo}, hereafter referred to as AnyTalk\textsubscript{\textit{MEMO}}. MEMO is built with a disentangled architecture that explicitly separates spatial and temporal processing similar to Hallo. Following the protocol of Section~\ref{sec:csf}, we freeze all of MEMO’s modules except its spatial residual layer in the denoising U-Net to apply $CsF$. We render the target 3D character under each active blendshape with zeroed-out audio embeddings to create a “no-motion” training set, which we use to fine-tune MEMO to obtain MEMO\textsubscript{\textit{CsF}}. We then perform our landmark-based blendshape optimization (Section~\ref{sec:optimization}) on the MEMO\textsubscript{\textit{CsF}} outputs to generate the 3D speech animation, thereby yielding AnyTalk\textsubscript{\textit{MEMO}}. Qualitative results are shown in Figure~\ref{fig:memo_extension}, demonstrating that lip synchronization aligns with the given phoneme. This extension confirms that our method, which comprises zeroed-out audio fine-tuning of spatial modules and subsequent landmark-driven blendshape optimization, can be straightforwardly integrated into different diffusion-based audio-driven talking-head video generation models.}

%% file: figure_tex/05_01_distillation.tex
\begin{figure}[t]
\centering
  \includegraphics[width=\linewidth]{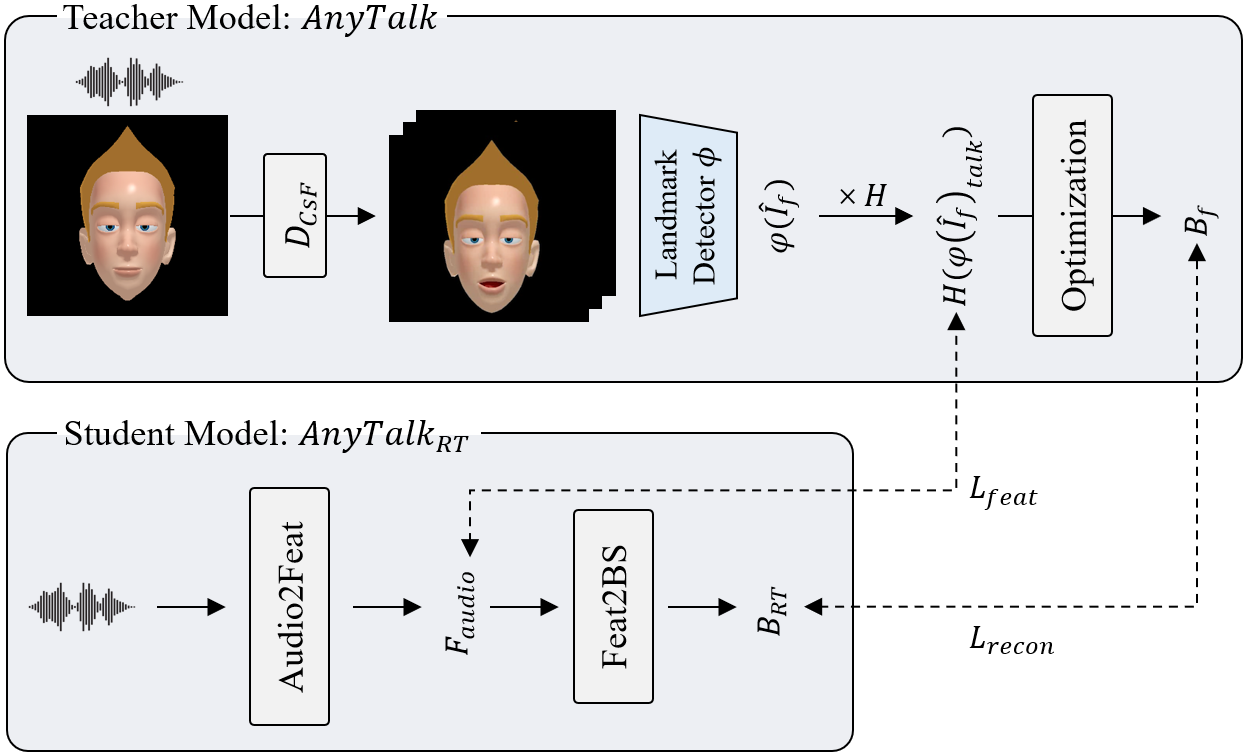}\vspace{-2mm}
  \caption{Distillation process from AnyTalk to $\text{AnyTalk}_{RT}$. Audio2Feat extracts features from the input audio, while Feat2BS predicts blendshape parameters from the extracted features.}
  \label{fig:distillation}
\end{figure}

%% file: sec/6_conclusion.tex
\section{Conclusion}
In this study, we presented an audio-driven speech animation method for arbitrary avatars, which does not require animation data—a critical advancement for real-world adoption that has not been previously explored. While previous speech-driven facial animation methods have primarily focused on supervised, character-specific scenarios, accumulating advances in extensive paired-data setting, our paper pioneers a fundamentally new direction: \fixq{animation-data-free} speech animation for arbitrary characters. We narrow the domain gap between the generated video and rendered images of the character using $CsF$ and zeroed-out audio embedding. Additionally, we proposed an optimization method that employs talk-invariant landmark based face alignment with landmark matching to create faithful speech animations. \fixq{While prioritizing 3D geometric stability and view-independence may lead to less pronounced lip movements compared to per-frame 2D-centric models, it ensures the structural consistency required for robust 3D lifting.} We believe that our method paves a way for \fixq{animation-data-free} audio-driven 3D animation, enabling its application to arbitrary characters regardless of mesh structure.

\paragraph{Limitations and Future Work.}
While our method produces promising results and enables significant advancements in speech animation for arbitrary characters, it also has challenges to address. Our method requires optimization during inference, which takes around 3.12 seconds for a frame. While we demonstrated the potential for real-time performance through distillation, there is a trade-off between generation quality and inference time. Addressing this trade-off would be crucial for future research. \fixq{A second limitation is the reliance on pre-defined blendshape parameters. Because our optimization process focuses on blendshape estimation, the input character must be rigged. Consequently, if a model is poorly rigged, for instance, lacking a blendshape for mouth opening, the animation may fail. Transitioning toward direct mesh animation could be a promising direction to address these constraints. Further more, our reliance on 2D landmarks may limit the capture of subtle facial details and high-frequency motions due to their sparse nature. While our experiments confirmed that incorporating dense photometric loss to address this introduces significant optimization ambiguity and overhead without qualitative gains, a possible future direction is to integrate hybrid dense-sparse representations, such as localized neural displacement maps, to recover these fine-grained dynamics.}

\fixq{Another} limitation is its dependence on frontal views. Although our final output is 3D speech animation, the optimization process relies on 3D landmarks. While landmarks have been widely adopted as essential part for 3D face capture and animation~\cite{zhou2018visemenet,feng2021learning,ling2022semantically,danvevcek2022emoca}, they still lack the expressiveness needed for rich detail. Simple attempts to incorporate multi-view information—such as using front, left, and right images for video generation or reconstructing the output with an off-the-shelf method~\cite{zhao2025hunyuan3d} failed to resolve this issue (see Section 3 of the supplementary material). Generating separate videos from different viewpoints produces inconsistent outputs due to the stochastic nature of diffusion, and off-the-shelf reconstruction methods do not produce 3D models that exactly match the target character. To overcome this single-view dependency, future research could explore the development of multi-view facial animation models~\cite{li2023efficient,li2024talkinggaussian,yun2025ffacenerf,zhou2023lc} that (1) control head and lip motion distinctly; (2) allow spatial layers to be tuned while freezing temporal layers independently; and (3) match the quality of state-of-the-art 2D video generation models. Developing such a method would preserve the animation quality of AnyTalk while enhancing multi-view consistency for the optimization process.

%% file: sec/99_supplementary.tex
\section{Content}
In this document, we provide additional experimental details, extended comparisons, and supplementary analyses that complement the main paper. The table of contents are as follows:

\begin{itemize}
    \item \textbf{Section~\ref{sec:supp-experiments}}: 
    Section~2.1 describes the dataset. Section~2.2 presents detailed implementations of baseline methods. Section~2.3 provides additional comparison results. Section~2.4 includes extended experiments, covering a sensitivity analysis of lip-sync metrics on stylized characters, evaluation excluding Character-specific Fine-tuning ($CsF$), and temporal stability analysis.
    
    \item \textbf{Section~\ref{sec:appendix-3d}}: 
    Section~3.1 analyzes inconsistent lip motion across viewpoints. Section~3.2 examines 3D reconstruction mismatches for unseen views. Section~3.3 discusses approaches for enhancing naturalness.
\end{itemize}

\section{Experiments Details} \label{sec:supp-experiments}

\subsection{Dataset}

\begin{figure}[h]
    \centering
    \vspace{-1mm}
    \includegraphics[width=\linewidth]{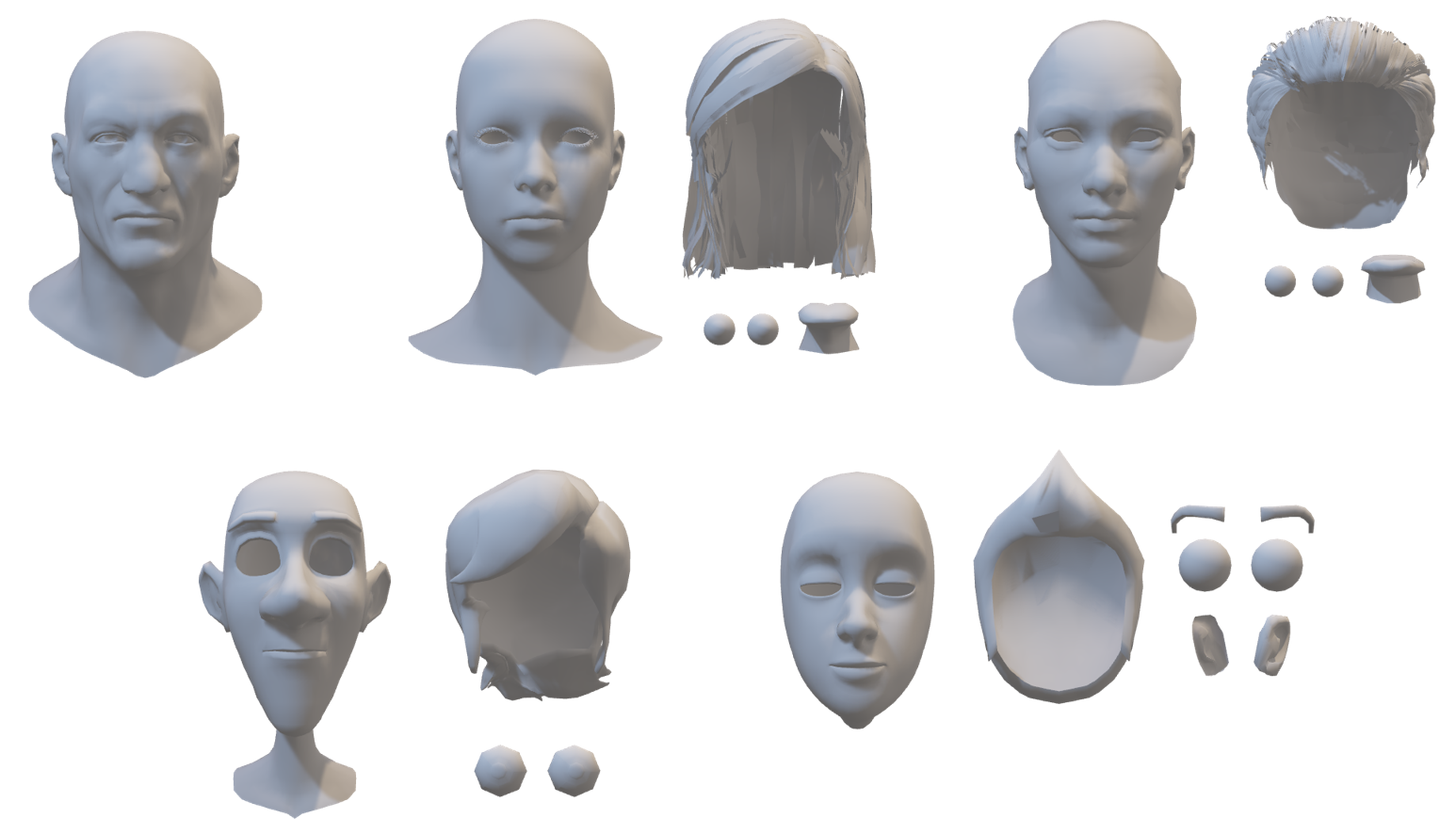}
    \caption{Mesh components of each character. From the top left to the bottom right: Victor, Emily, VMan, Malcolm, Morphy.}
    \label{fig:all-meshes}
\end{figure}

We used five characters, each with distinct mesh structures, styles, and blendshape configurations. Victor consists of a single mesh that includes all facial components. In contrast, Emily, VMan, Malcolm, and Morphy have their facial elements divided into separate meshes. Specifically, Emily and VMan include the face, hair, left/right eyeballs, and tongue as separate meshes. Malcolm includes the face, hair, and left/right eyeballs, while Morphy includes the face, hair, eyebrows, left/right eyeballs, and left/right ears. The components of each character are shown in Figure~\ref{fig:all-meshes}.

\subsection{Baseline Comparison}

\subsubsection{ScanTalk}

To generate audio-driven speech animations using ScanTalk, a single uniform face mesh was used for Victor, whereas separated face meshes were utilized for the other characters. The resulting animations were subsequently combined with the remaining meshes to produce the final results. Furthermore, to match the scale of each mesh with the VOCASET-like meshes used in ScanTalk training, the center points and scales of the meshes were calculated and adjusted to align with VOCASET~\cite{cudeiro2019capture}.

\subsubsection{DiffSpeaker}

DiffSpeaker was trained on VOCASET and performs inference using the same dataset. We generated audio-driven speech animation using VOCASET as input. The original mesh includes eyeballs, but to perform retargeting with NFR, the eyeballs were removed from the original mesh using the FLAME mask before inputting it into the model.

\subsubsection{CodeTalker}
\fixq{CodeTalker~\cite{xing2023codetalker} is a supervised model that encodes facial actions using a discrete motion prior, generating 3D facial motion with expressive lip and facial movements. Similar to DiffSpeaker, CodeTalker was trained on the BIWI and VOCASET datasets. Therefore, we generated audio-driven speech animation using VOCASET as input and applied the FLAME mask to remove the eyeballs from the original mesh. Then, we performed retargeting using NFR. While the VOCA mesh maintained its structure without distortion during speech animation generation, the retargeting process introduced mesh distortion, leading to reduced lip synchronization quality.}

\subsubsection{NFR}
\begin{figure}[t]
    \centering
    \includegraphics[width=0.5\linewidth]{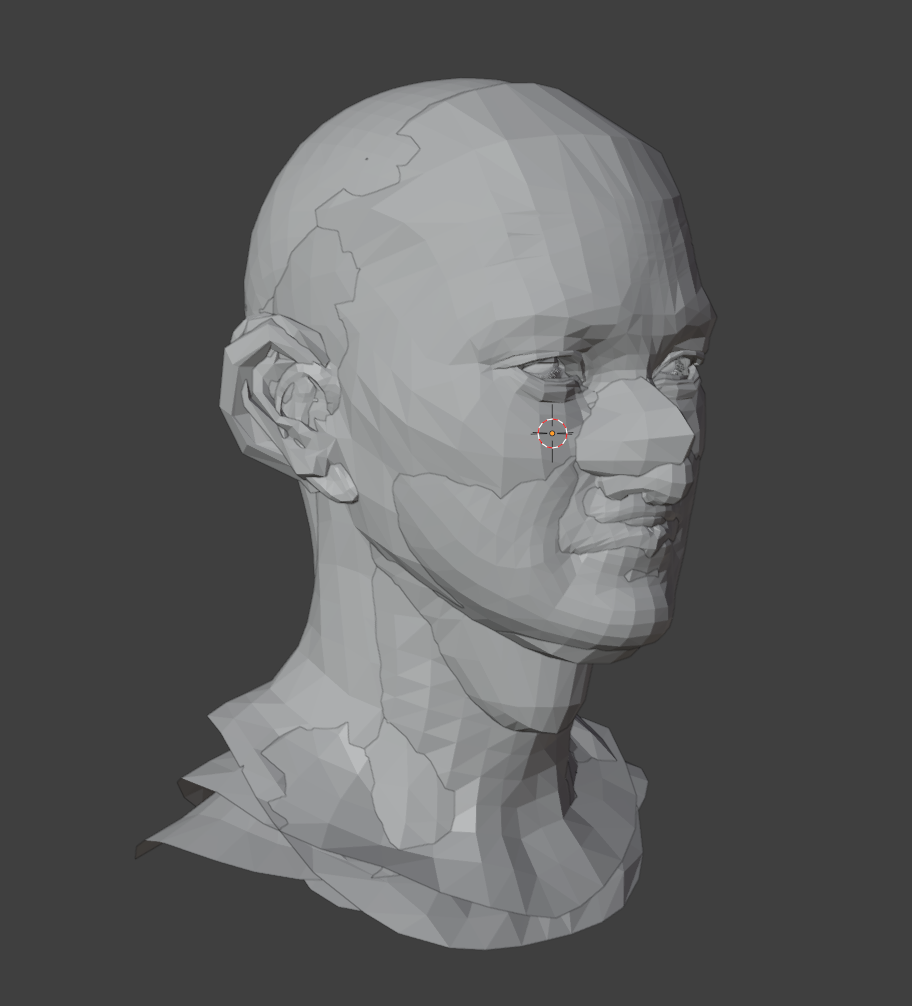}
    \caption{Mesh alignment for baseline methods in Blender.}
    \label{fig:mesh-alignment}
\end{figure}

When using NFR, its paper~\cite{qin2023neural} recommends using a face mesh without the mouth, eyes, nose sockets, and eyeballs to avoid wrong deformations. However, Victor includes all these elements in a single structure, and as there is no definition of which element each vertex corresponds to, separating them is not feasible. While removing vertices corresponding to specific elements is possible using 3D software, this approach was deemed unsuitable for comparisons involving arbitrary avatars. Our method is designed to generate speech animations without requiring manual preprocessing, regardless of the mesh structure or blendshape configuration. Therefore, Victor, including the mouth, eyes, nose sockets, and eyeballs, was used as input. For the other characters—Emily, VMan, Malcolm, and Morphy—separated face meshes were used as input. Additionally, the align.blend file provided by NFR was used to match the scale of the meshes with those used during NFR training. This adjustment is illustrated in Figure~\ref{fig:mesh-alignment}. For cases where separated face meshes were used, the output was combined with the remaining meshes (e.g., hair, eyeballs) to produce the final result.

\input{figure_tex/99_02_comparison}

\subsection{Additional Comparison} \label{sec:sup-comparison}

We compared our approach with additional method, Choi et al. + Hallo~\cite{choi2024deep, xu2024hallo}, using all 5 characters and 30 audio clips, resulting in a total of 150 speech animations. Qualitative comparison results are presented in Figure~\ref{fig:comparison-supplementary}, while quantitative evaluation results are shown in Table~\ref{tab:quant_baseline2}. As shown in Figure~\ref{fig:comparison-supplementary}, the lip movements generated by our method more precisely follow the given sentence. In contrast, Choi et al. + Hallo produced only subtle movements that did not align with the phonemes. Table~\ref{tab:quant_baseline2} shows that our method performed \fixq{significantly better compared to Choi et al.} across both metrics.

\subsubsection{Choi et al.}
Choi et al.~\cite{choi2024deep} is a semi-supervised video-based retargeting method that estimates blendshape parameters for 3D stylized characters from a source human performance video using an image translation-based approach. For the source video, we used the video of the first identity from MEAD dataset for both training and inference. Although the method does not require a paired dataset for each source video and the animation frames of the target character, the target frames, along with the corresponding blendshape parameters, must exist and match the range of motion (ROM) of the source video, which does not exist. To address this, we activated blendshape parameters randomly and stitched them together to create target animation frames. 
During inference, we used Hallo to generate the source video. We obtained the code from the authors to run the method.

\begin{table}[t]
    \centering
    \caption{Quantitative results comparing our method with baselines. Best denoted in bold.}
    \label{tab:quant_baseline2}
    \scalebox{0.92}{
    {\renewcommand{\arraystretch}{1.2}
        \begin{tabular}{lcc}
            \hline
            \textbf{Methods} & LSE-D$\downarrow$ & LSE-C$\uparrow$ \\ \hline
            Ours & \textbf{11.304} & \textbf{3.155} \\ \hline
            Choi et al. + Hallo & 16.571 & 0.285 \\ \hline \vspace{-3mm}
        \end{tabular}
    }
    }
    \vspace{-3mm}
\end{table}

\subsection{\fixq{Additional Experiments}} \label{sec:sup-addexperiment}
\subsubsection{\fixq{Sensitivity Analysis of Lip-Sync Metrics on Stylization}}
\fixq{To evaluate whether Lip-Sync Error (LSE) metrics, originally trained on natural human faces, remain reliable for stylized characters, we conducted a sensitivity analysis. We compared the performance of stylized characters (Morphy and Malcolm) against photorealistic non-stylized characters (Emily, VMan, and Victor).} \fixq{The results, summarized in Table~\ref{tab:sup-addexp-style}, indicate that while non-stylized characters achieve slightly better scores, the performance gap is not statistically significant. Specifically, LSE-D increased by only 5.4\%, and LSE-C decreased by 4.6\% when moving from photorealistic to stylized domains. This suggests that, despite the domain shift in appearance, the metrics effectively capture the underlying lip dynamics, which in our stylized assets still closely follow human-like phonetic movements. Thus, LSE-D and LSE-C remain valid comparative indicators for stylized character animation.}

\begin{table}[t]
    \centering
    \caption{Comparison of lip-sync metrics between stylized and non-stylized characters. The marginal difference suggests that SyncNet-based embeddings remain robust to artistic stylization.}
    \label{tab:sup-addexp-style}
    \scalebox{0.95}{
    {\renewcommand{\arraystretch}{1.2}
        \begin{tabular}{lcc}
            \hline
            \textbf{Characters} & \textbf{LSE-D}$\downarrow$ & \textbf{LSE-C}$\uparrow$ \\ \hline
            Stylized (Morphy, Malcolm) & 11.65 & 3.07 \\ \hline
            Non-Stylized (Emily, VMan, Victor) & 11.05 & 3.22 \\ \hline \vspace{-3mm}
        \end{tabular}
    }
    }
\end{table}

\subsubsection{\fixq{Evaluation Excluding Personalization}} \label{sec:sup-addexp-personalize}
\fixq{To evaluate the performance of our method under different usage regimes, we conducted experiments distinguishing between personalized and generic inference settings. While our full pipeline employs $CsF$ to adapt the underlying 2D generator to a target identity, baseline methods such as ScanTalk, DiffSpeaker, and CodeTalker operate in a generic setting, relying on post-hoc alignment or retargeting without identity-specific optimization. To ensure a fair comparison when quantifying the contribution of our core architecture independent of this personalization, we include a non-personalized variant of our method (ours w/o $CsF$) in our evaluation. All experiments follow the same settings as our ablation study.}

\fixq{As reported in Table~\ref{tab:sup-addexp-without-tuning}, even without character-specific tuning, our base model significantly outperforms the retargeting-based baselines (DiffSpeaker + NFR and CodeTalker + NFR) across all metrics and achieves highly competitive lip-sync accuracy compared to ScanTalk. Furthermore, \fixq{ours} w/o $CsF$ achieves the highest lip-sync confidence among all evaluated methods, demonstrating the robustness of our underlying motion synthesis framework. The application of $CsF$ in our full pipeline further refines the lip-sync precision, yielding the best overall LSE-D while maintaining strong confidence.}

\begin{table}[t]
    \centering
    \caption{Quantitative comparison with baselines and our non-personalized variant (w/o $CsF$) to evaluate the impact of character-specific tuning.}
    \label{tab:sup-addexp-without-tuning}
    \scalebox{0.95}{
    {\renewcommand{\arraystretch}{1.2}
        \begin{tabular}{lcc}
            \hline
            \textbf{Methods} & \textbf{LSE-D}$\downarrow$ & \textbf{LSE-C}$\uparrow$ \\ \hline
            ScanTalk & \underline{10.787} & 2.955 \\ \hline
            DiffSpeaker + NFR & 13.393 & 0.559 \\ \hline
            CodeTalker + NFR & 13.392 & 0.550 \\ \hline
            Ours w/o $CsF$ & 11.207 & \textbf{3.451} \\ \hline
            Ours & \textbf{10.695} & \underline{3.397} \\ \hline \vspace{-3mm}
        \end{tabular}
    }
    }
    \vspace{-3mm}
\end{table}

\subsubsection{\fixq{Temporal Stability of AnyTalk and $\text{AnyTalk}_{RT}$}} \label{sec:sup-addexp-temporal}

\fixq{To quantitatively assess the temporal stability of our primary model, AnyTalk, compared to its real-time variant, $\text{AnyTalk}_{RT}$, we measured the temporal smoothness of the generated facial animations. Specifically, we computed the average jerk (the rate of change of acceleration) of facial landmarks across consecutive generated frames, where lower jerk values indicate smoother, less jittery motion. The facial landmarks used for this evaluation are extracted using XPose~\cite{yang2025x}. We evaluated this metric under two configurations: the overall jerk across all detected facial landmarks, and the talk-related jerk, which isolates the landmarks specifically associated with the mouth and lip regions. As detailed in Table~\ref{tab:sup-anytalk-anytalkRT-jitter}, AnyTalk exhibits lower jerk values in both the full facial configuration and the localized talk-related configuration. These results quantitatively demonstrate that while $\text{AnyTalk}_{RT}$ successfully distilled the architecture for real-time inference, the full AnyTalk model (acting as the teacher network) maintains a slight advantage in preserving temporal stability and producing naturally smooth facial dynamics.}

\begin{table}[t]
    \centering
    \caption{Quantitative comparison of temporal stability of AnyTalk and AnyTalk\_RT.}
    \label{tab:sup-anytalk-anytalkRT-jitter}
    \scalebox{0.95}{
    {\renewcommand{\arraystretch}{1.2}
        \begin{tabular}{lcc}
            \hline
            \textbf{Methods} & \textbf{Jerk}$\downarrow$ & \textbf{Talk-related jerk}$\downarrow$ \\ \hline
            AnyTalk & \textbf{27.297} & \textbf{14.419} \\ \hline
            $\text{AnyTalk}_{RT}$ & 27.603 & 14.462 \\ \hline \vspace{-3mm}
        \end{tabular}
    }
    }
\end{table}

\section{Multi-view Generation and 3D Reconstruction Analysis}
\label{sec:appendix-3d}
To further investigate the limitations of our single-view optimization pipeline and naive solutions cannot be adopted to address this issue, we conducted two supplementary experiments: (1) generating talking-head videos from multiple viewpoints, left (${-45^\circ}$), front (${0^\circ}$), and right (${+45^\circ}$) using Hallo~\cite{xu2024hallo}, and (2) reconstructing the resulting frames into 3D meshes using Hunyuan3D-2~\cite{zhao2025hunyuan3d}. The results confirm the challenges discussed in the main text.

\begin{figure}[t!]
    \centering
    \includegraphics[width=0.95\linewidth]{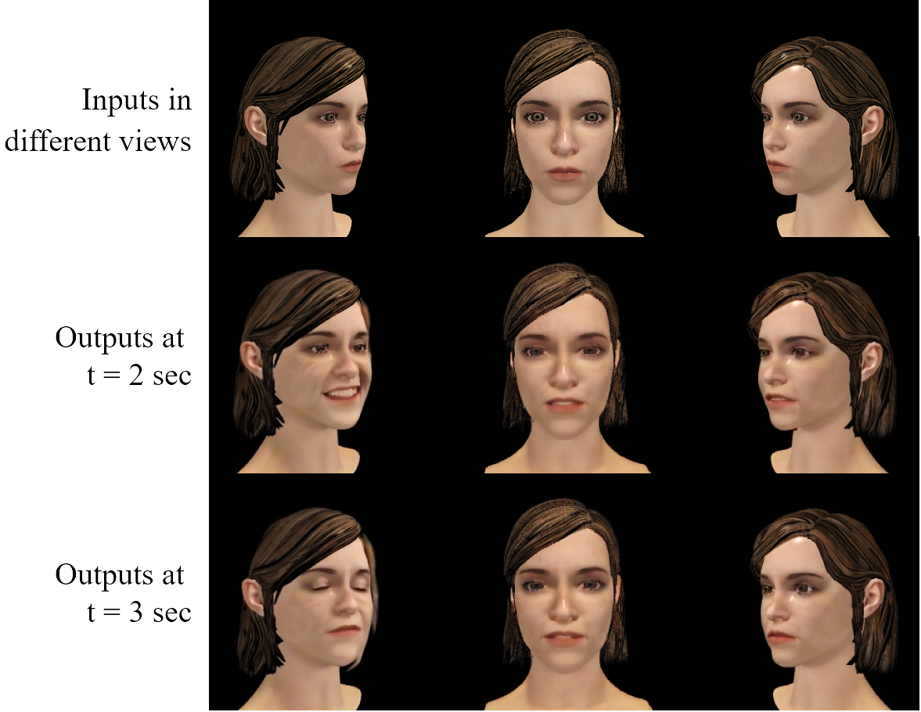}
    \vspace{-2mm}
    \caption{Multi-view generation results for the same audio segment at time $t$. The top row shows the input character renders for left, frontal, and right viewpoints. The middle and bottom rows display the corresponding generated video frames at identical timestamps, illustrating inconsistent lip motions and facial expressions across views. }
    \label{fig:video-differentviews}
\end{figure}

\begin{figure}[t!]
    \centering
    \includegraphics[width=0.76\linewidth]{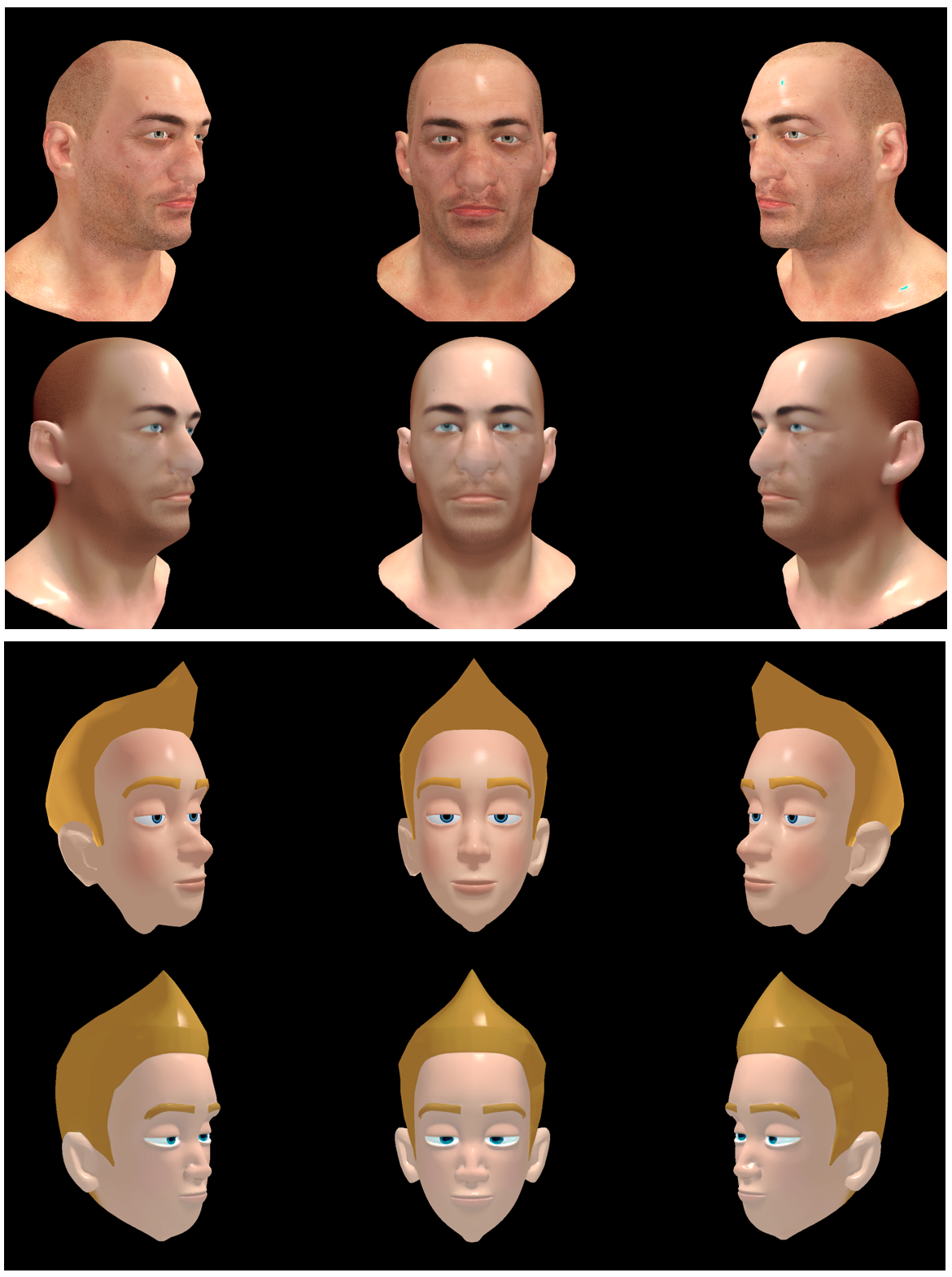}
    \vspace{-1mm}
    \caption{Comparison between original meshes (first and third rows) and 3D reconstructed meshes (second and fourth rows) generated by Hunyuan3D-2. While the reconstructed characters resemble the originals, they lack precise pixel-wise alignment, making them unsuitable for direct motion guidance.}
    \label{fig:3drecon}
    \vspace{-3mm}
\end{figure}

\subsection{Inconsistent Lip Motion Across Viewpoints}
We generated three separate videos of the same input audio: one each for the frontal, left, and right views. As shown in Figure~\ref{fig:video-differentviews}, although the overall head pose approximately matches the intended viewpoints, the lip motion and facial expressions vary noticeably between views. This inconsistency arises from the stochastic sampling inherent in diffusion models, \fixq{which leads to non-deterministic outputs, even for the same audio content. Consequently, this lack of inter-view coherence introduces conflicting supervision signals, which prevents the 3D representation from converging to a sharp, geometrically consistent state.} Such variability undermines the reliability of a multi-view optimization strategy that assumes consistent motion priors across generated videos.

\subsection{3D Reconstruction Mismatch for Unseen Views}
Next, we applied a state-of-the-art 3D reconstruction method~\cite{zhao2025hunyuan3d} to the generated multi-view frames, aiming to recover a mesh that matches the original character for 3D guidance. Figure~\ref{fig:3drecon} illustrates that the reconstructed geometry often deviates from the source character’s shape and proportions particularly for viewpoints of left and right. Artifacts such as distorted jawlines and misaligned landmarks indicate that generic reconstruction methods struggle to faithfully reproduce the subtle details of our target meshes when given diffusion-generated images as input.

These results together demonstrate that naively combining diffusion generation in multi-view or using off-the-shelf reconstruction is insufficient to resolve the single-view dependency of our current system. Future work should therefore explore integrated approaches for multi-view consistency in talking-head diffusion models or end-to-end 3D speech animation frameworks that jointly optimize for both view synthesis and geometric fidelity.

\subsection{Enhancing Naturalness}
Because our method utilizes blendshapes of an arbitrary character, adding random expressions, such as random eye blinks, can be easily achieved as follows:
\begin{align}
w =
\begin{cases}
\dfrac{t-s}{d_i}, & s \le t < s+ d_i/2,\\[6pt]
\dfrac{d_i - t + s }{d_i}, & s+d_i/2 \le t < s+d_i,\\[4pt]
0, & \text{otherwise.}
\end{cases}
\end{align}
Here, $s$ denotes the start time of the expression, $t$ is the current time, and $d_i$ is the duration of the expression. By simply applying random values to $s$ and $d_i$, the character can easily exhibit dynamic expressions. Alternatively, we can directly apply a constant expression throughout the sequence by assigning $w = 1$. Examples demonstrating random blinking and constant expressions are provided in the supplementary video.

%% file: figure_tex/99_02_comparison.tex
\begin{figure*}[ht]
\centering \vspace{-2mm}
  \includegraphics[width=\linewidth]{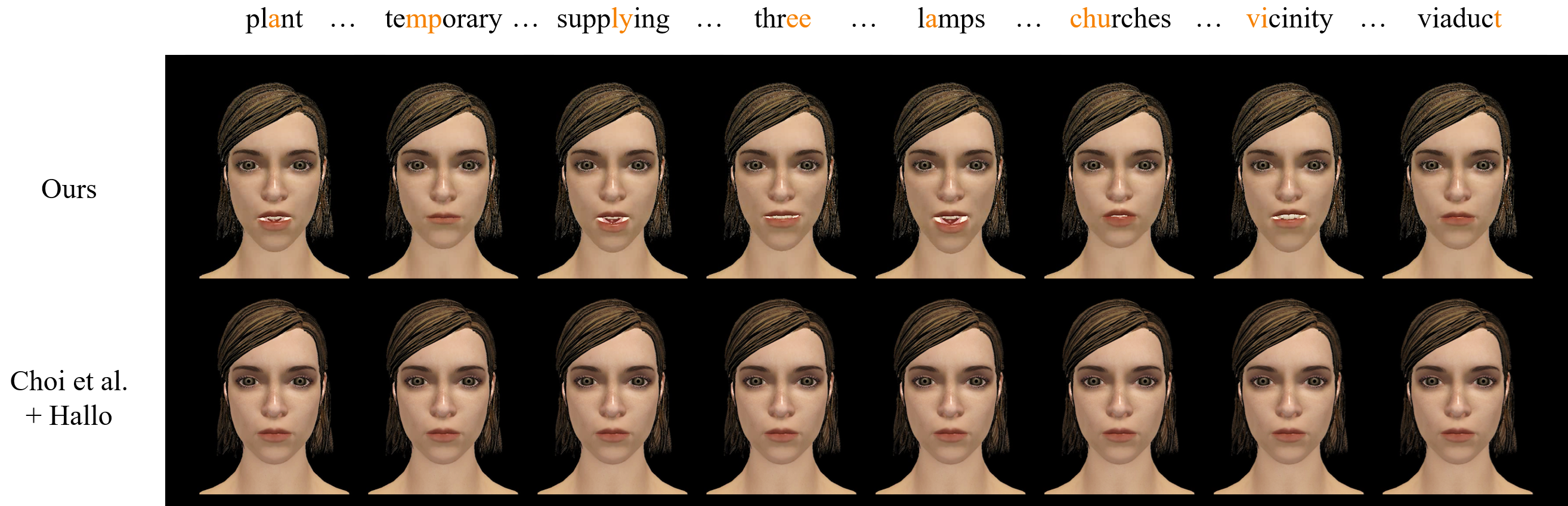} \vspace{-3mm}
  \caption{\fixq{Comparison with additional baselines on Emily. Each letter and its corresponding frame for comparison is presented. Ours best follows the given audio while Choi et al. + Hallo show only subtle movements.}}
  \label{fig:comparison-supplementary}
\end{figure*}